\documentclass[10pt]{article}

\usepackage{authblk}

\usepackage{graphicx} 
\usepackage[scriptsize,nooneline,hang]{caption}
\usepackage[hang,nooneline,scriptsize]{subfigure}

\usepackage{amsmath,amsfonts,bbm,dsfont,mathrsfs}

\newcommand{\dd}{{\mathrm{d}}}

\begin{document}

\title{Thermodynamics of a rotating black hole in minimal five-dimensional gauged supergravity}

\author[1,2]{Saskia Grunau}
\author[1]{ Hendrik Neumann}
\affil[1]{Institut f\"ur Physik, Universit\"at Oldenburg, D--26111 Oldenburg, Germany}
\affil[2]{Niels Bohr Institute, University of Copenhagen, DK--2100 Copenhagen, Denmark}
	
\maketitle

\begin{abstract}
In this article we study the thermodynamics of a general non-extremal rotating black hole in minimal five-dimensional gauged supergravity. We analyse the entropy-temperature diagram and the free energy. Additionally we consider the thermodynamic stability by calculating the specific heat, the isothermal moment of inertia tensor and the adiabatic compressibility.
\end{abstract}

\section{Introduction}

In 1997, the famous anti-de Sitter/ conformal field theory (AdS/CFT) correspondence connected quantum field theory to gravity \cite{PD_Maldacena:1997re}. It was proposed that compactifications of string theory on anti-de Sitter space are related to a conformal field theory. Therefore, asymptotically anti-de Sitter black holes are an interesting phenomenon to study.

Considering black hole thermodynamics, new behaviour occurs in anti-de Sitter spacetimes. The asymptotically flat Schwarzschild black hole is not thermodynamically stable, its heat capacity is negative and it will evaporate emitting Hawking radiation. In the Schwarzschild-AdS spacetime a second branch of stable solutions emerges. Then there are two possible phases for a given temperature, small unstable and larger stable black holes \cite{PD_Hawking:1982dh}. If the AdS black hole is rotating or carries a charge, there a two black hole phases with positive heat capacity. So there is a first order phase transition from small to large black holes, which interestingly resembles the liquid/gas transition of an ideal Van der Waals gas. The Reissner-Nordstr\"om AdS spacetime and its thermodynamical properties were analysed in \cite{PD_Chamblin:1999tk, PD_Chamblin:1999hg} and the Kerr AdS spacetime was studied in \cite{PD_Hawking:1998kw}. Also five-dimensional AdS black holes with two independent rotation parameters were considered \cite{PD_Hawking:1998kw}.
In the Kerr Newman AdS spacetime, the effect of charge and additionally rotation on the thermodynamics was studied in \cite{PD_Caldarelli:1999xj}. Also higher dimensions have been investigated. In $d\geq 6$ dimensions a so called reentrant phase transition occurs in the Kerr-AdS spacetime, describing a large/small/large black hole transition \cite{PD_Altamirano:2013ane}. In the case of $d=6$ multiple rotating Kerr AdS black holes, even analogues of solid/liquid and solid/liquid/gas phase transition were found \cite{PD_Altamirano:2013uqa}.
\\

The positivity of the specific heat is an important factor regarding the thermodynamic stability of a black hole. Furthermore, the moment of inertia must be positive. In higher dimensions multiple angular momenta are possible, turning the moment of inertia into a tensor. Then all eigenvalues of the moment of inertia tensor have to be positive \cite{PD_Monteiro:2009tc, PD_Dolan:2013yca}.
However, in asymptotically AdS spacetimes, an additional condition has to be taken into account. Since the cosmological constant can be interpreted as a pressure with an effective volume as the conjugated thermodynamic variable, this gives rise to a new quantity, the adiabatic compressibility \cite{PD_Dolan:2013dga}. This compressibility has to be positive for thermodynamic stability \cite{PD_Dolan:2014lea}.
\\

In this paper we will study the general non-extremal rotating black hole in minimal five-dimensional gauged supergravity \cite{PD_Chong:2005hr}. In addition to two independent angular momenta, this solution also carries charge. We will first give a review of the metric and the thermodynamic quantities. Then possible phase transitions will be analysed using the state equation $T(S)$ and the Gibbs free energy. In the last section we will consider the thermodynamic stability by calculating the specific heat, the moment of inertia tensor and the adiabatic compressibility.

\section{The metric and its thermodynamic quantities}

First we will give a brief overview of the general non-extremal black hole in five dimensional gauged supergravity and its properties. The metric was found by Chong, Cveti\u{c}, L\"u and Pope \cite{PD_Chong:2005hr} and is given by
\begin{align}
 \dd s^2 = &-\frac{\Delta_\theta [(1+g^2r^2)\rho^2\dd t+2q\nu ]\dd t}{\Xi_a\Xi_b\rho^2} + \frac{2q\nu\omega}{\rho^2} + \frac{f}{\rho^4} \left( \frac{\Delta_\theta\dd t}{\Xi_a\Xi_b} -\omega \right)^2 + \frac{\rho^2 \dd r^2}{\Delta_r} \nonumber\\ 
 &+ \frac{\rho^2 \dd \theta^2}{\Delta_\theta} 
 + \frac{r^2+a^2}{\Xi_a}\sin^2\theta \, \dd\phi^2 + \frac{r^2+b^2}{\Xi_b}\cos^2\theta \, \dd\psi^2 \, ,
\end{align}
with the gauge potential
\begin{equation}
 A=\frac{\sqrt{3}q}{\rho^2}\left( \frac{\Delta_\theta \dd t}{\Xi_a\Xi_b} -\omega \right) \, .
\end{equation}
Here the metric is presented in asymptotically static Boyer-Lindquist coordinates, the metric in an asymptotically rotating frame can be found in \cite{PD_Aliev:2008yk}. The solution is characterised by the mass, the charge and two independent rotation parameters. The parameter $g$ is related to the cosmological constant $\Lambda$ by $g^2=-\frac{\Lambda}{4}$. 

The metric functions are
\begin{align}
 \nu &= b\sin^2\theta \, \dd\phi+a\cos^2\theta \, \dd\psi \, , \nonumber\\
 \omega &= a\sin^2\theta \, \frac{\dd\phi}{\Xi_a}+b\cos^2\theta \, \frac{\dd\psi}{\Xi_b} \, , \nonumber\\
 \Xi_a &= 1-a^2g^2 \, , \nonumber\\
 \Xi_b &= 1-b^2g^2 \, , \nonumber\\
 \Delta_\theta &= 1- a^2g^2\cos^2\theta - b^2g^2\sin^2\theta \, , \nonumber\\
 \Delta_r &= \frac{(r^2+a^2)(r^2+b^2)(1+g^2r^2)+q^2+2abq}{r^2}-2m \, , \nonumber\\
 \rho^2 &=  r^2+a^2\cos^2\theta +b^2\sin^2\theta \, , \nonumber\\
 f &= 2m\rho^2-q^2+2abqg^2\rho^2 \, .
\end{align}
The horizons of the black hole are solutions of $\Delta_r=0$. We define $r=r_+$ to be the largest positive root, i.e. the outer event horizon. The angular velocities at the event horizon are
\begin{align}
 \Omega_a &= \frac{a(r_+^2+b^2)(1+g^2r_+^2)+bq}{(r_+^2+a^2)(r_+^2+b^2)+abq} \, , \nonumber\\
 \Omega_b &= \frac{b(r_+^2+a^2)(1+g^2r_+^2)+aq}{(r_+^2+a^2)(r_+^2+b^2)+abq} \, .
 \label{eqn:PD_omega}
\end{align}
With the Killing vector field $\xi = \partial_t + \Omega_a\partial_\phi +  \Omega_b\partial_\psi$ we can now calculate the surface gravity 
\begin{equation}
\kappa^2=\left. -\frac{1}{2} (\nabla_\mu\xi_\nu) (\nabla^\mu\xi^\nu) \right|_{r=r_+}
\end{equation}
and thus the Hawking temperature
\begin{equation}
 T = \frac{\kappa}{2\pi} = \frac{r_+^4[1+g^2(2r_+^2+a^2+b^2)]-(ab+q)^2}{2\pi r_+ [(r_+^2+a^2)(r_+^2+b^2)+abq]} \, .
 \label{eqn:PD_temperature}
\end{equation}
The entropy $S=\frac{A_{\rm H}}{4}$ can be found from the area of the horizon
\begin{equation}
 A_{\rm H} = \int_0^{2\pi}\! \int_0^{2\pi}\! \int_0^{\frac{\pi}{2}} \! \sqrt{\det (g_{\rm H})} \, \dd\theta\dd\phi\dd\psi \, ,
\end{equation}
where $g_{\rm H}$ is the metric tensor at the horizon. Hence the entropy is
\begin{equation}
 S=\frac{\pi^2[(r_+^2+a^2)(r_+^2+b^2)+abq]}{2\Xi_a\Xi_b r_+} \, . \label{eqn:PD_entropy}
\end{equation}
The two angular momenta can be determined using Komar integrals
\begin{equation}
 J=\frac{1}{16\pi}\int_{S^3}\!\ast \dd \zeta \, ,
\end{equation}
where $\zeta = \partial_\phi$ or  $\zeta =\partial_\psi$ depending on the desired angular momentum.In terms of the metric components we get\\
\begin{align}
 J_a =  \frac{1}{16\pi} \int_{S^3}\! \frac{g_{\theta\theta}\, \dd\theta\dd\phi\dd\psi}{\sqrt{-\det (g)}} &\left[ (g_{\phi\psi}^2-g_{\phi\phi}g_{\psi\psi})\partial_r g_{t\phi} + (g_{t\phi}g_{\psi\psi}-g_{t\psi}g_{\phi\psi})\partial_r g_{\phi\phi} 
 + (g_{t\psi}g_{\phi\phi}-g_{t\phi}g_{\phi\psi})\partial_r g_{\phi\psi}\right] \, , \nonumber\\
 J_b = \frac{1}{16\pi} \int_{S^3}\! \frac{g_{\theta\theta}\, \dd\theta\dd\phi\dd\psi}{\sqrt{-\det (g)}} &\left[ (g_{\phi\psi}^2-g_{\phi\phi}g_{\psi\psi})\partial_r g_{t\psi} + (g_{t\psi}g_{\phi\phi}-g_{t\phi}g_{\phi\psi})\partial_r g_{\psi\psi} 
 + (g_{t\phi}g_{\psi\psi}-g_{t\psi}g_{\phi\psi})\partial_r g_{\phi\psi}\right]  \, . 
\end{align}
The angular momenta are
\begin{align}
 J_a &= \frac{\pi [2am+qb(1+a^2g^2)]}{4\Xi_a^2\Xi_b} \, , \nonumber \\
 J_b &= \frac{\pi [2bm+qa(1+b^2g^2)]}{4\Xi_a\Xi_b^2} \, . \label{eqn:PD_angularmomenta}
\end{align}
Furthermore, the electrostatic potential at the horizon $\Phi=\left. \xi^\mu A_\mu \right|_{r=r_+}$ is given by
\begin{equation}
 \Phi = \frac{\sqrt{3}qr_+^2}{(r_+^2+a^2)(r_+^2+b^2)+abq} \, .
\end{equation}
The charge can be found using a Gaussian integral
\begin{equation}
 Q = \frac{1}{16\pi}\int_{S^3}\! \left(\ast F - F \wedge \frac{A}{3}\right)  \quad {\rm with} \quad F=dA \, , \label{eqn:PD_charge}
\end{equation}
therefore
\begin{equation}
 Q = \frac{\sqrt{3}\pi q}{4\Xi_a\Xi_b} \, .
\end{equation}
The conserved mass follows from the integration of the first law of thermodynamics
\begin{equation}
 M = \frac{m\pi (2\Xi_a+2\Xi_b-\Xi_a\Xi_b) + 2\pi qabg^2(\Xi_a+\Xi_b)}{4\Xi_a^2\Xi_b^2} \, . \label{eqn:PD_mass}
\end{equation}
Note that the value of $m$ in the equations \eqref{eqn:PD_angularmomenta} and  \eqref{eqn:PD_mass} is determined by $\Delta_r(r=r_+)=0$.

The cosmological constant can be related to a pressure \cite{PD_Cvetic:2010jb}
\begin{equation}
 P=-\frac{D-2}{16\pi}\Lambda = \frac{(D-2)(D-1)}{16\pi}g^2 \, ,
\end{equation}
so for this spacetime in five dimensions we have
\begin{equation}
 P=\frac{3g^2}{4\pi} \, .
\end{equation}
Then there is also a conjugate thermodynamic variable $V$, which can be seen as an effective volume inside the horizon \cite{PD_Cvetic:2010jb}
\begin{equation}
 V=\frac{\pi^2}{2\Xi_a\Xi_b} \left[(r_+^2+a^2)(r_+^2+b^2)+\frac{2}{3}abq \right]+\frac{2\pi}{3}\left(aJ_a+bJ_b\right) \, .
\end{equation}
In the presence of a cosmological constant, the first law of thermodynamics is changed to \cite{PD_Cvetic:2010jb}
\begin{equation}
 \dd M= T\dd S + \Omega_a\dd J_a + \Omega_b\dd J_b + \Phi\dd Q + V\dd P \, ,
\end{equation}
where $M$ can be seen as the total gravitational enthalpy.

\section{State equation}
\label{sec:PD_state-equation}
In this section we analyse the state equation $T(S, J_a, J_b, Q)$ keeping the charge and the two angular momenta fixed. Also the pressure $P$ will be held constant. While an analytic formula for $T$ as a function of the thermodynamic variables cannot be given, it is possible to obtain the result numerically.
First one chooses $g$ (this will also set the pressure $P$), $Q$, $J_a$ and $J_b$. Then the equations \eqref{eqn:PD_angularmomenta} and \eqref{eqn:PD_charge} are solved for the $a$, $b$ and $q$. Now $r_+$ is the last remaining parameter and $T(r_+)$ versus $S(r_+)$ can be plotted.

\begin{figure}[!h]
	\centering
	\subfigure[$g=0$ and $J_b=0$]{
		\includegraphics[width=0.44\textwidth]{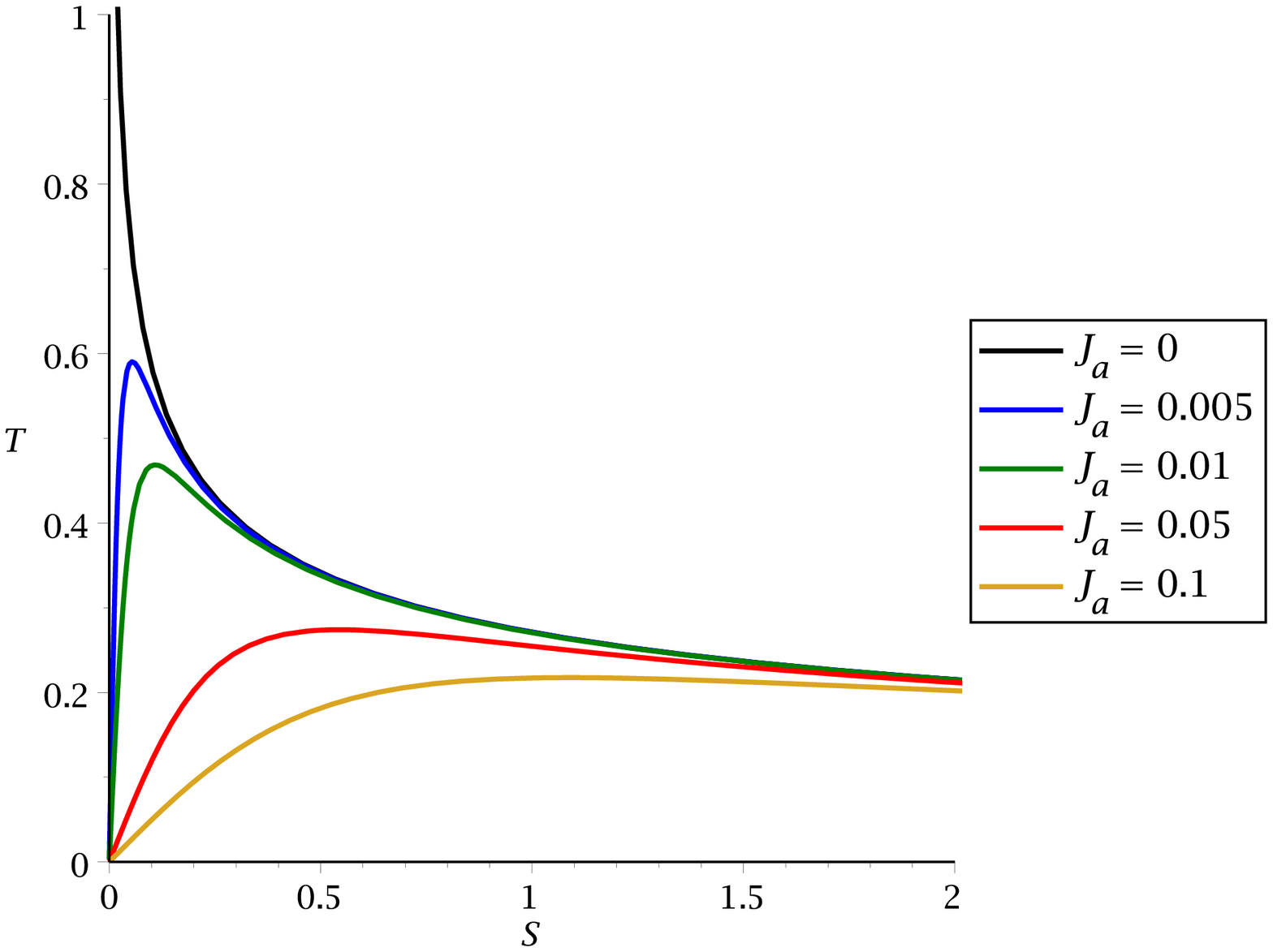}
	}
	\subfigure[$g=0$ and $J_b=0.01$]{
		\includegraphics[width=0.44\textwidth]{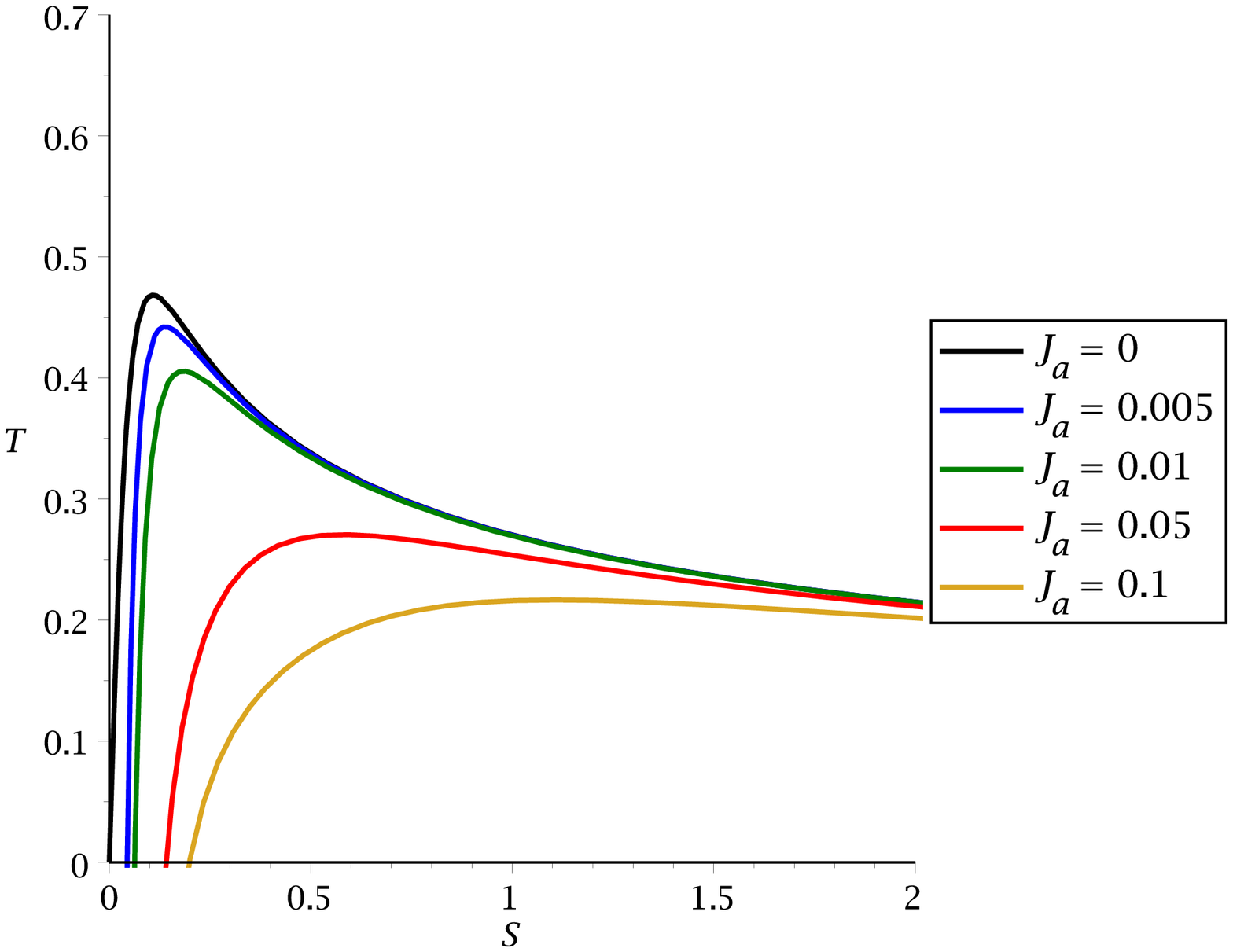}
	}
	\subfigure[$g=1$ and $J_b=0$]{
		\includegraphics[width=0.44\textwidth]{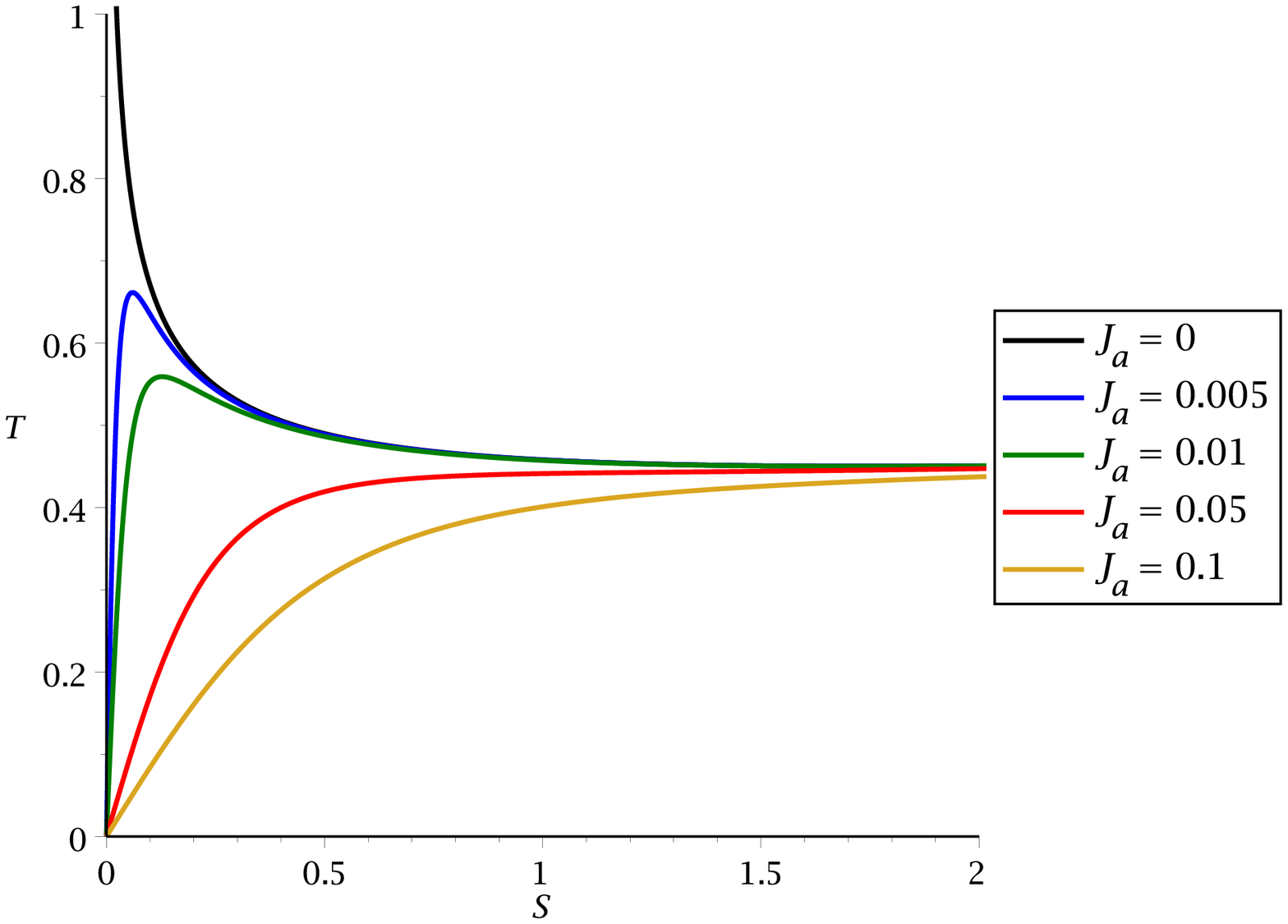}
	}
	\subfigure[$g=1$ and $J_b=0.005$]{
		\includegraphics[width=0.44\textwidth]{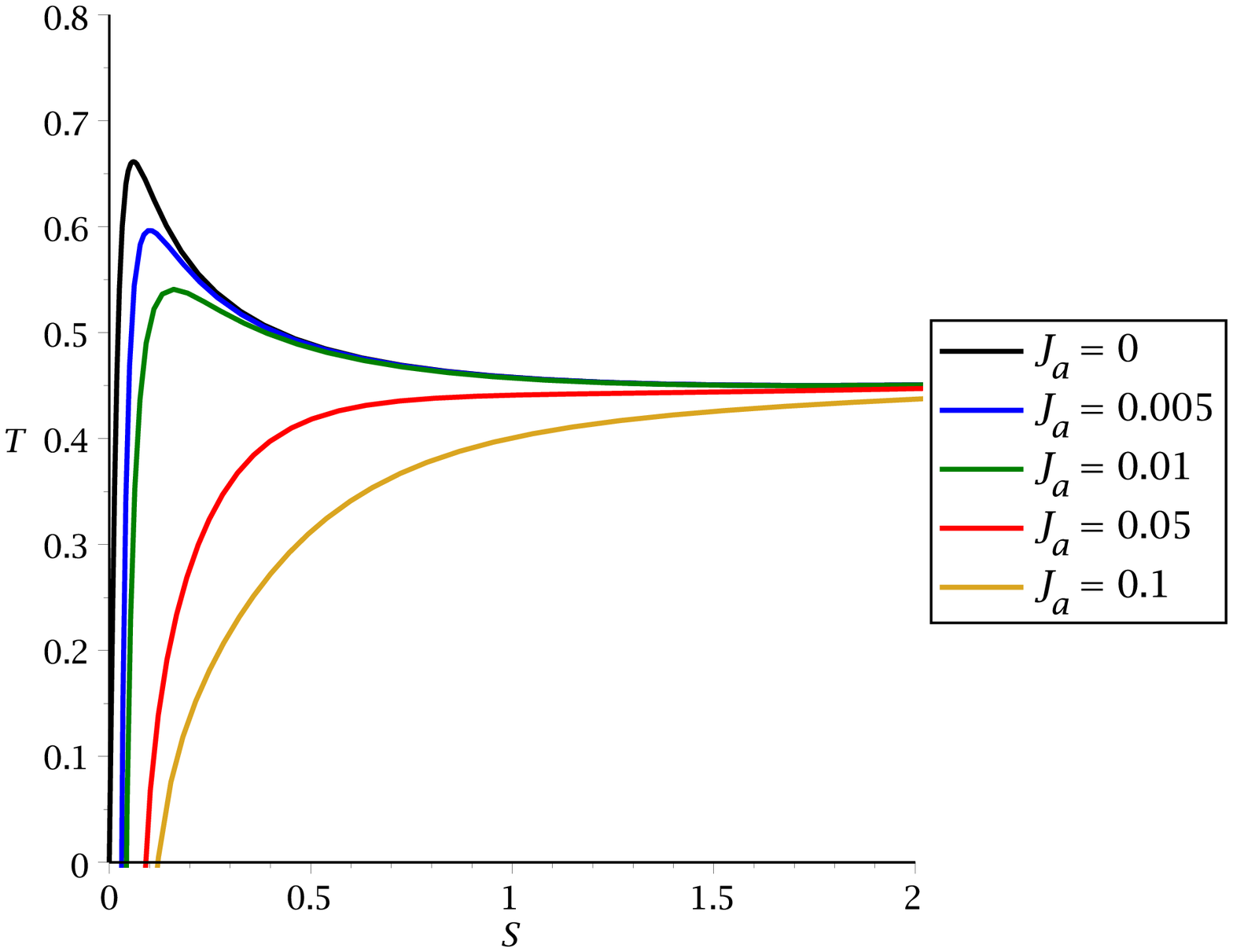}
	}
	\caption{Temperature versus entropy for  $Q=0$.}
	\label{pic:PD_TS-Q0}
\end{figure}

Figure \ref{pic:PD_TS-Q0} shows several numerically obtained plots for $Q=0$. In this case the metric reduces to the uncharged Kerr black hole of \cite{PD_Hawking:1998kw} with two independent rotation parameters. $T(S)$ curves for $g=0$ and $J_b=0$ are shown in figure \ref{pic:PD_TS-Q0}(a), here we see the usual Schwarzschild ($J_a=0$) and Kerr ($J_a\neq 0$) behaviour. In the Schwarzschild case $T$ diverges at $S=0$ and then decreases with increasing $S$. Therefore the specific heat 
\begin{equation}
 C=T\frac{\partial S}{\partial T}
\end{equation}
is always negative, i.e. in this case the black hole is unstable.
For $J_a\neq 0$, all curves start at $T=S=0$ and the temperature increases up to a maximum, here the specific heat is positive, which is one of the conditions for stability (see also section \ref{sec:PD_stability}). After the maximum the black hole is unstable. 

If $J_b\neq 0$ (figure \ref{pic:PD_TS-Q0}(b)), only the curve for $J_a=0$ starts at $T=S=0$, for $J_a\neq 0$ zero temperature is reached at a positive value of $S$. In the limit $S\rightarrow \infty$ the temperature converges to zero.

This behaviour changes once the cosmological constant is added to the spacetime ($g\neq 0$). $T(S)$ plots for $g=1$ are depicted in figure \ref{pic:PD_TS-Q0}(c) and (d). There the temperature no longer converges to zero, but increases for $S\rightarrow \infty$. Depending on the values of the two angular momenta we find a local maximum followed by a local minimum. So there is a branch of small stable black holes, followed by intermediate unstable black holes and finally a branch of large stable black holes. With increasing angular momenta the extrema merge into a point of inflection, leaving a unique stable black hole phase. This indicates a second order phase transition similar to the Van der Waals liquid/gas transition.
For $J_a=J_b=0$ the temperature diverges for $S=0$. In this case there is an unstable phase up to a minimum of $T$ and a phase of large stable black holes.
\\

\begin{figure}[!h]
 \centering
 \subfigure[$g=0$ and $J_b=0$]{
  \includegraphics[width=0.44\textwidth]{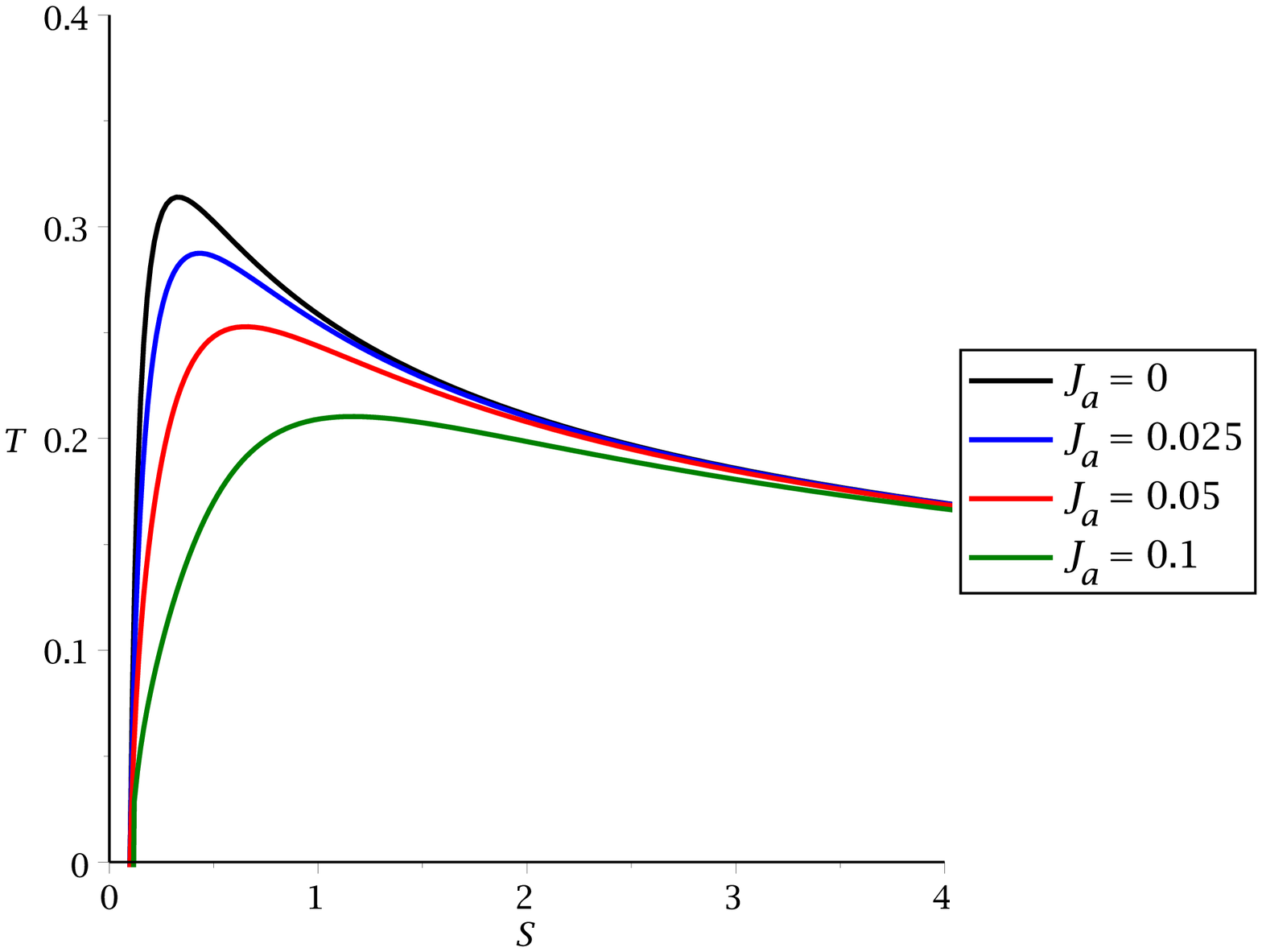}
 }
 \subfigure[$g=0$ and $J_b=0.025$]{
  \includegraphics[width=0.44\textwidth]{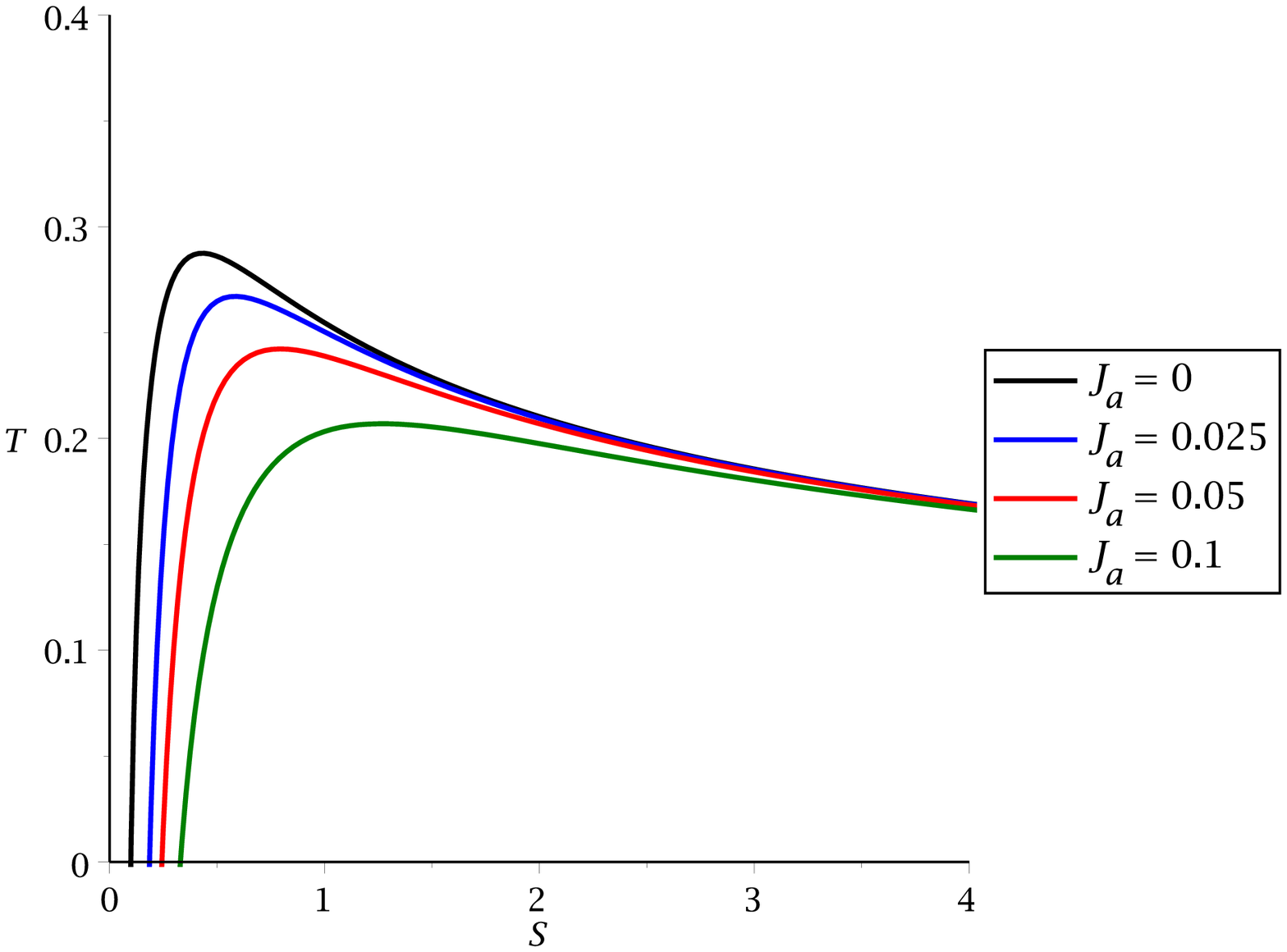}
 }
 \subfigure[$g=1$ and $J_b=0$]{
  \includegraphics[width=0.44\textwidth]{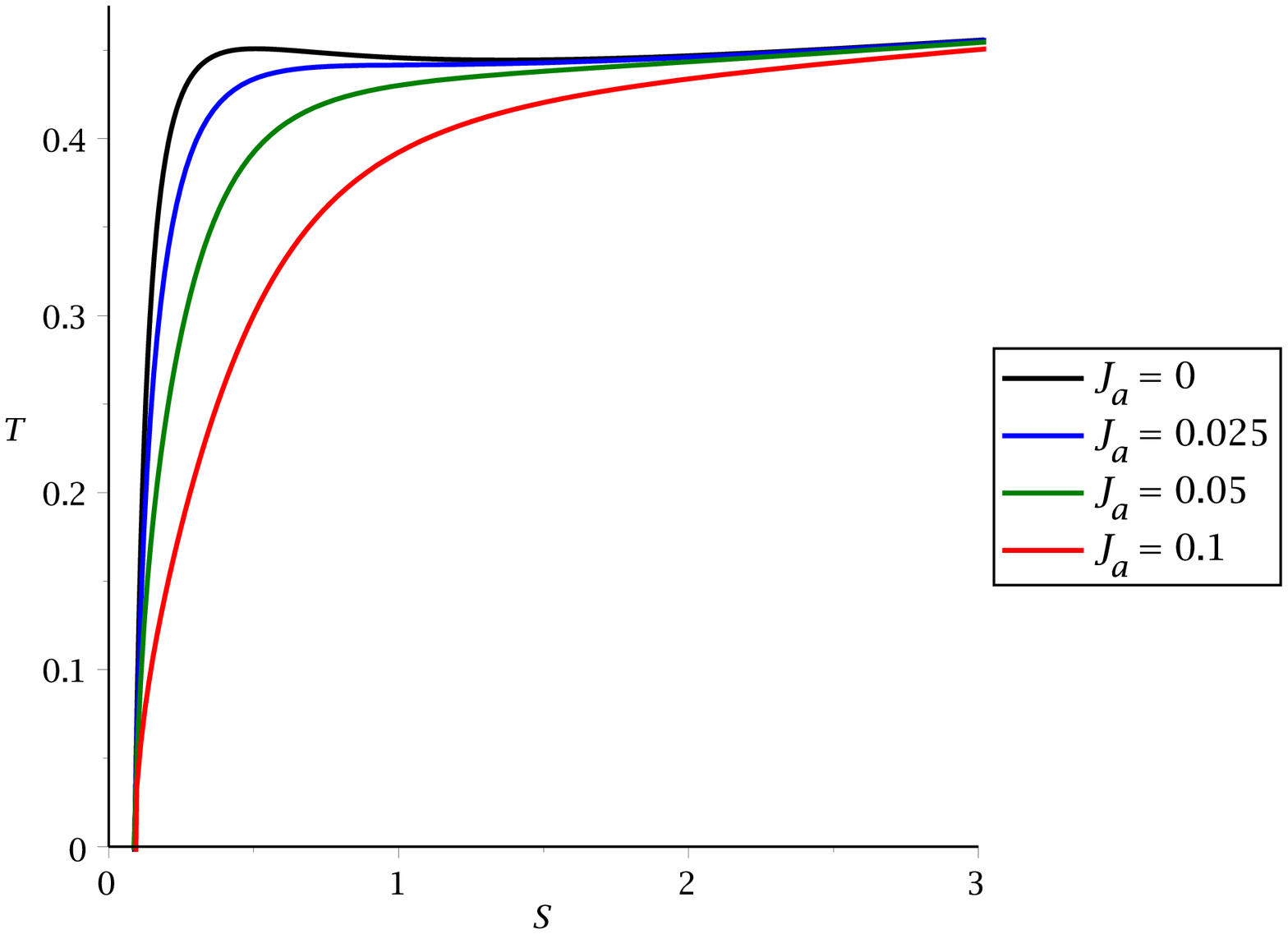}
 }
 \subfigure[$g=1$ and $J_b=0.025$]{
  \includegraphics[width=0.44\textwidth]{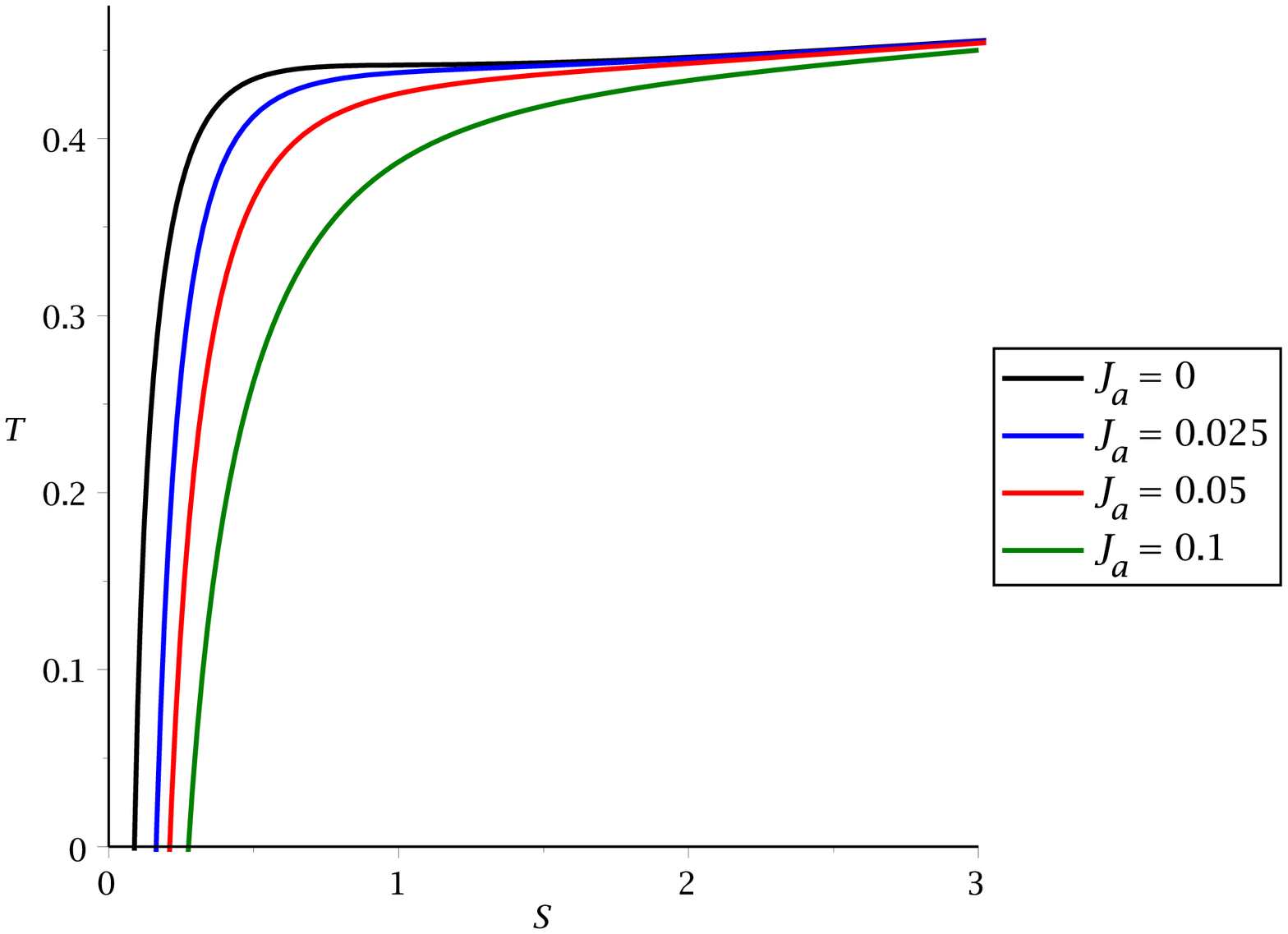}
 }
 \caption{Temperature versus entropy for  $Q=0.1$.}
 \label{pic:PD_TS-Q01}
\end{figure}

If the charge is nonzero, $T=0$ is reached for all values of $J_a$ and $J_b$ including $J_a=J_b=0$, see figure \ref{pic:PD_TS-Q01}. Also here a local maximum and a local minimum exist for certain values of the angular momenta and the charge. If $J_a$, $J_b$ and $Q$ exceed critical values, the local extrema transform into a point of inflection.

\begin{figure}[!h]
	\centering
	\subfigure[$Q=0.2$]{
		\includegraphics[width=0.43\textwidth]{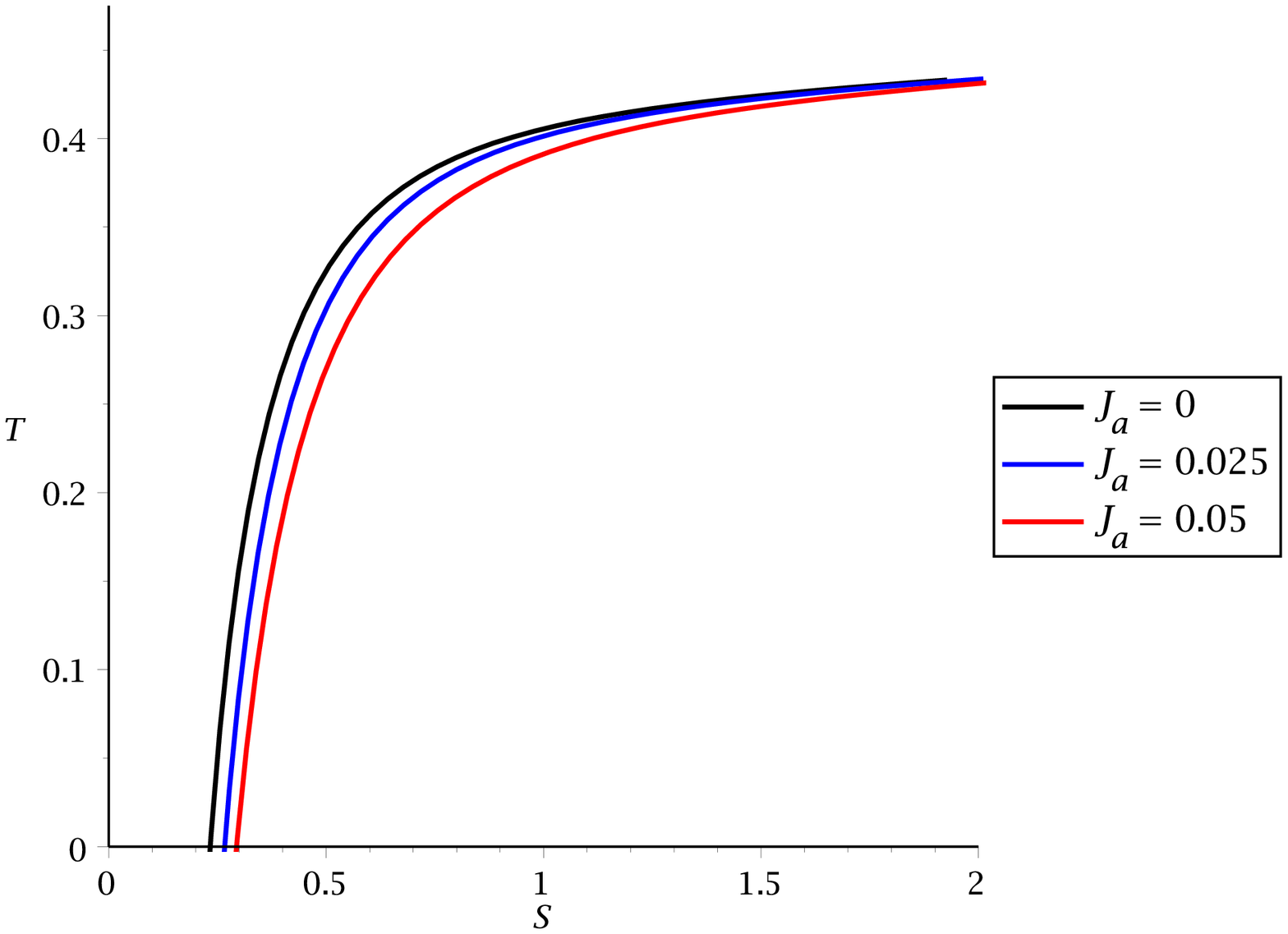}
	}
	\subfigure[$Q=-0.2$]{
		\includegraphics[width=0.43\textwidth]{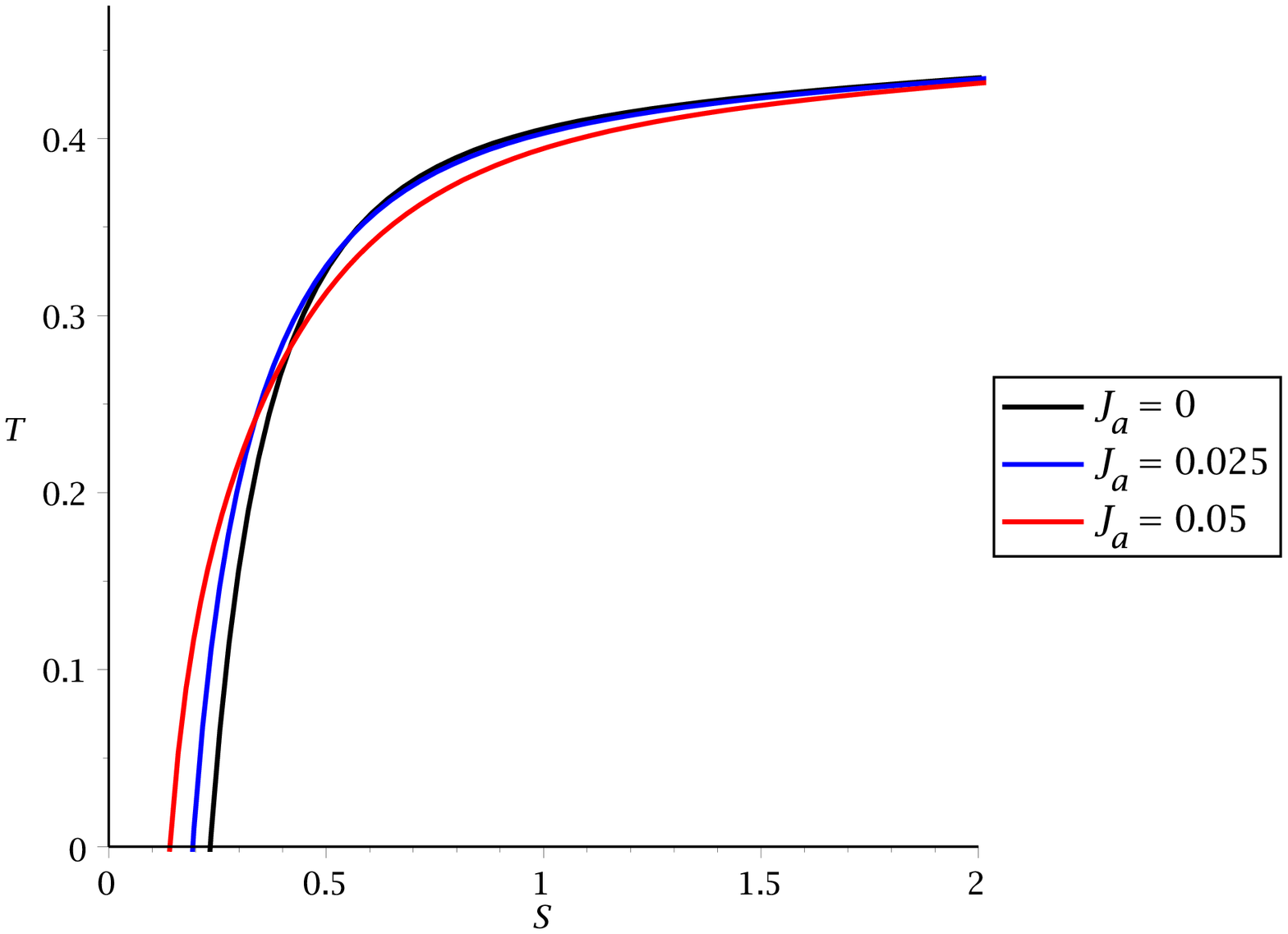}
	}
	\caption{Temperature versus entropy for $g=1$ and $J_b=0.025$. The different behaviour of the curves in the case of large negative charge may result from the influence of the Chern-Simons term.}
	\label{pic:PD_negcharge}
\end{figure}

For larger negative charges the curves change, see figure \ref{pic:PD_negcharge}. This effect may be due to the Chern-Simons term. However, there is no influence on the phase transitions or thermodynamic behaviour.

\begin{figure}[!h]
 \centering
 \includegraphics[width=0.5\textwidth]{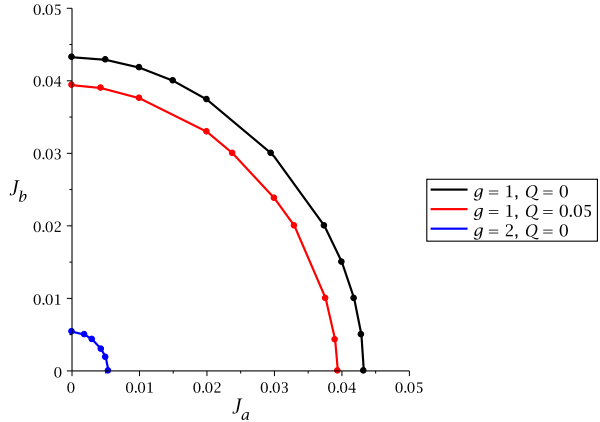}
 \caption{Second-order critical points in the $J_a$-$J_b$-plane.}
 \label{pic:PD_criticalpoints}
\end{figure}

The critical values can be found by calculating the derivative $\frac{\partial T}{\partial S}$. At the critical points a local minimum of $\frac{\partial T}{\partial S}$ coincides with $\frac{\partial T}{\partial S}=0$; we find these points numerically using a bisection method. Figure \ref{pic:PD_criticalpoints} shows the critical values of $J_a$ and $J_b$ for fixed $g$ and $Q$. Inside the curve there are two stable phases, a small black hole and a large black hole connected by a first order phase transition. Outside the curve a unique stable black hole phase exists. Here the properties of the rotating black hole in minimal five-dimensional gauged supergravity are similar to those of the Kerr Newman AdS spacetime \cite{PD_Caldarelli:1999xj}. The curve of critical points shrinks towards smaller angular momenta with increasing $Q$ and $g$, so the region with two phases becomes smaller until it vanishes completely. If the curve was extended to the whole $J_a$-$J_b$-plane, it would form a circle.

\section{Free energy}
\begin{figure}[!h]
	\centering
	\subfigure[$g=0$, $Q=0$ and $J_b=0$]{
		\includegraphics[width=0.46\textwidth]{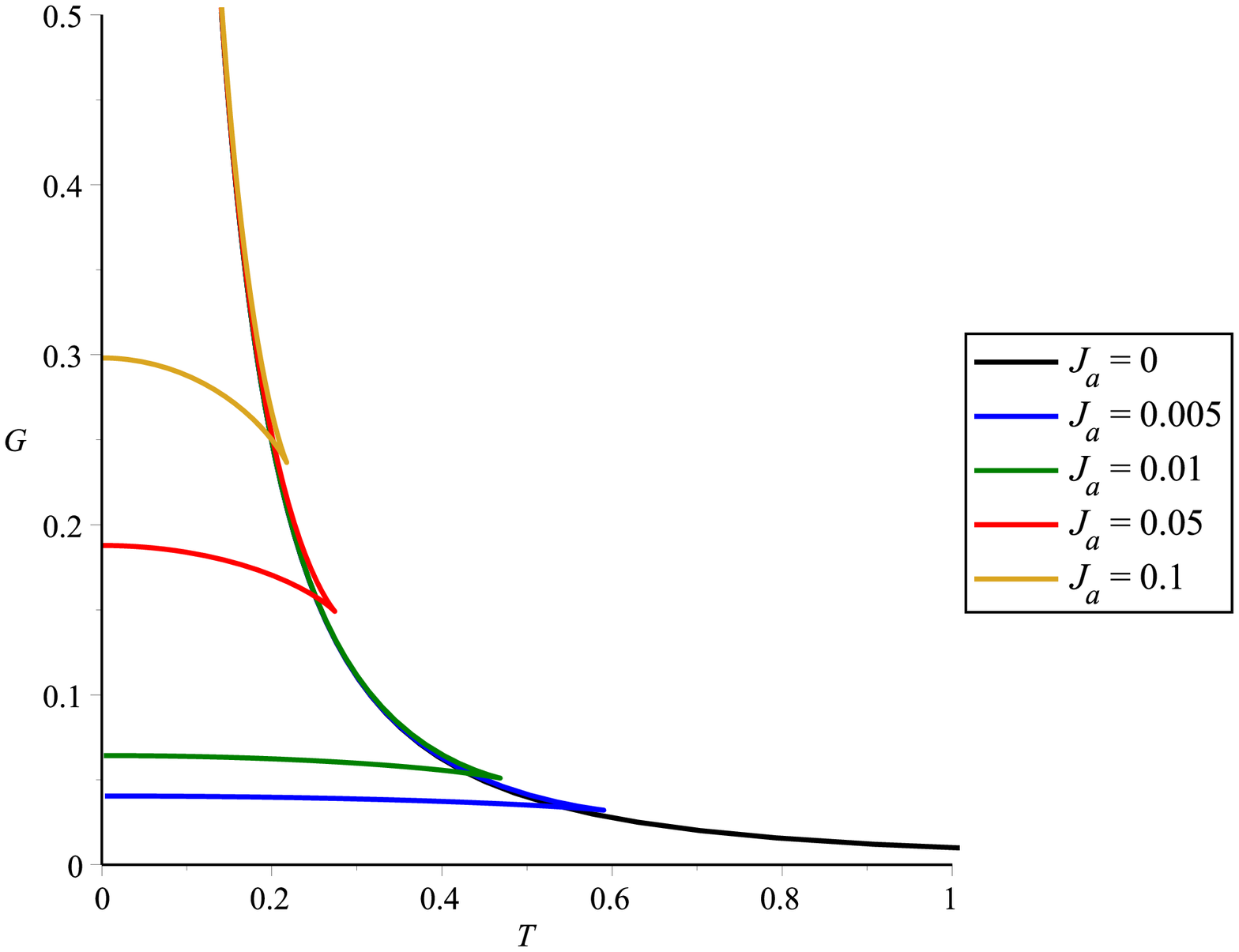}
	}
	\subfigure[$g=1$, $Q=0$ and $J_b=0$]{
		\includegraphics[width=0.46\textwidth]{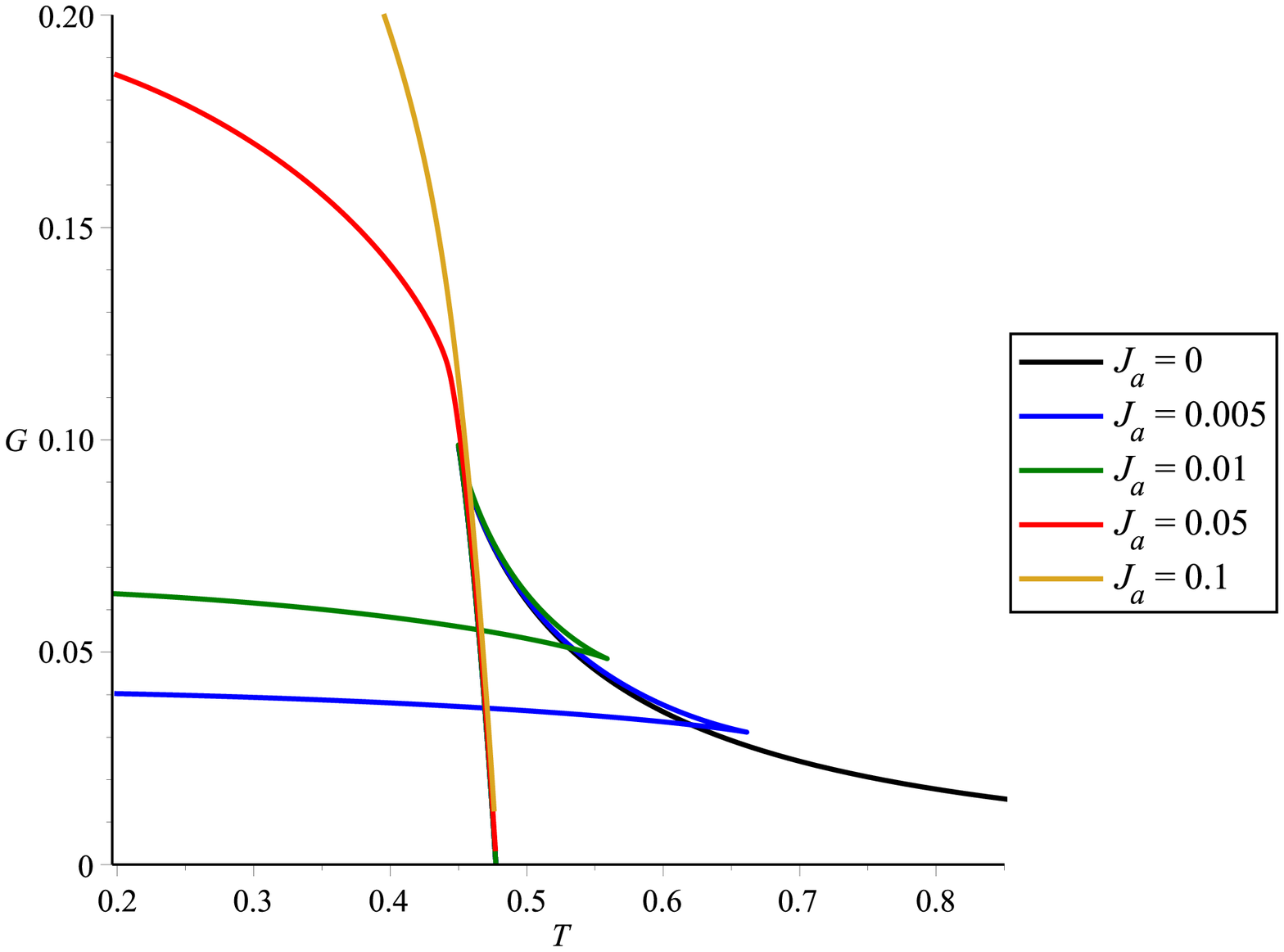}
	}

	\subfigure[$g=0$, $Q=0$ and $J_b=0.01$]{
		\includegraphics[width=0.46\textwidth]{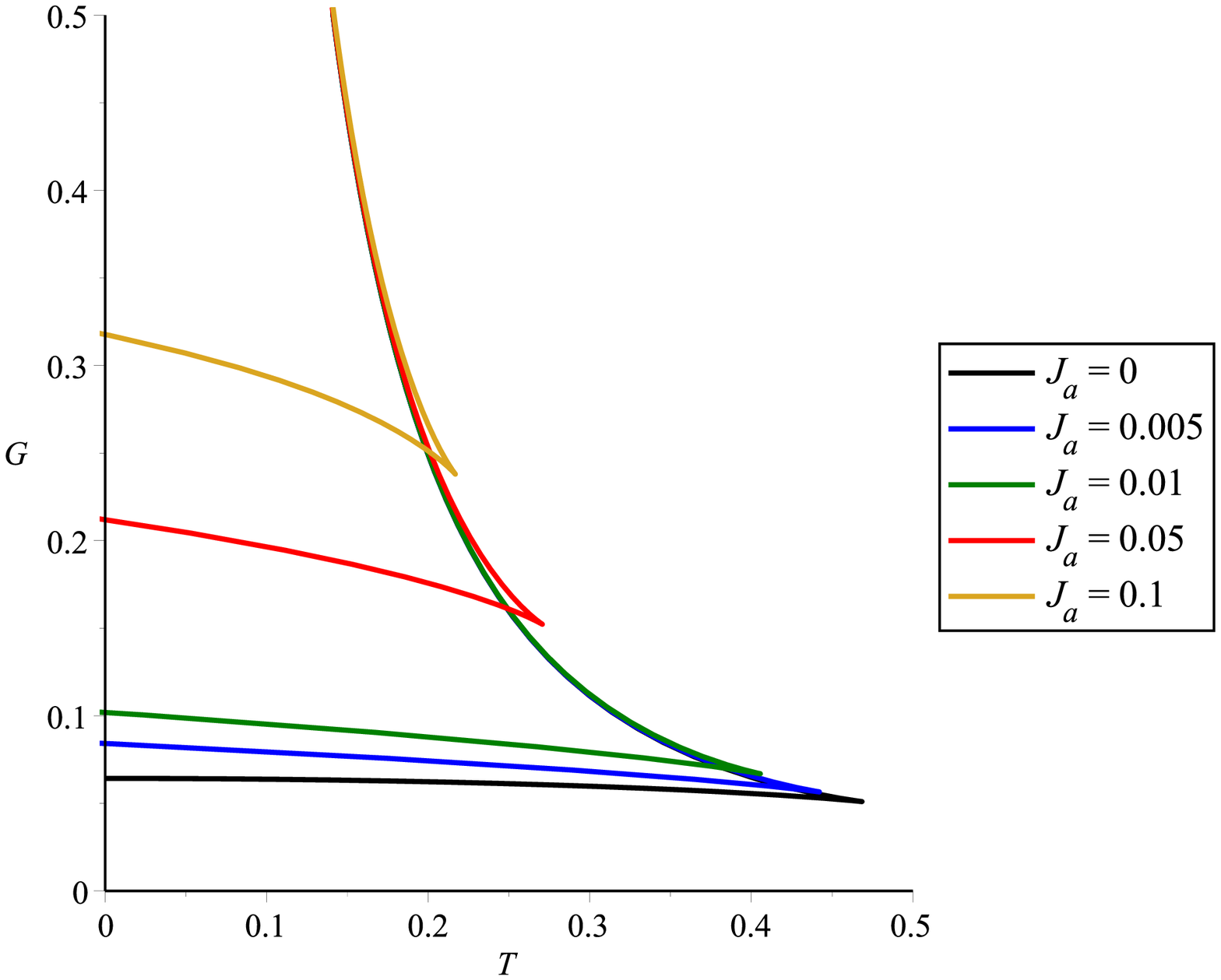}
	}
	\subfigure[$g=1$, $Q=0$ and $J_b=0.005$]{
		\includegraphics[width=0.46\textwidth]{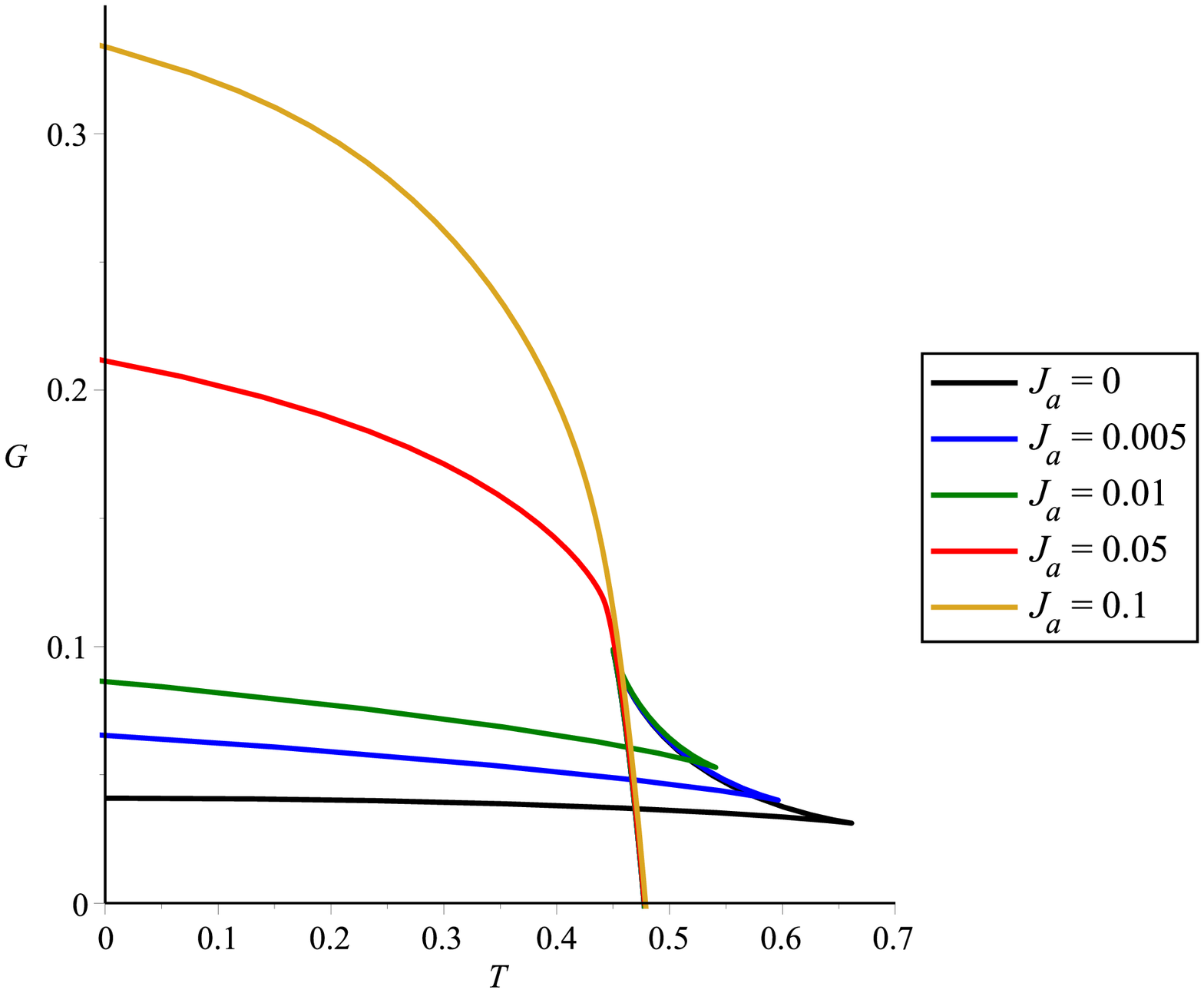}
	}

	\subfigure[$g=0$, $Q=0.1$ and $J_b=0.025$]{
		\includegraphics[width=0.46\textwidth]{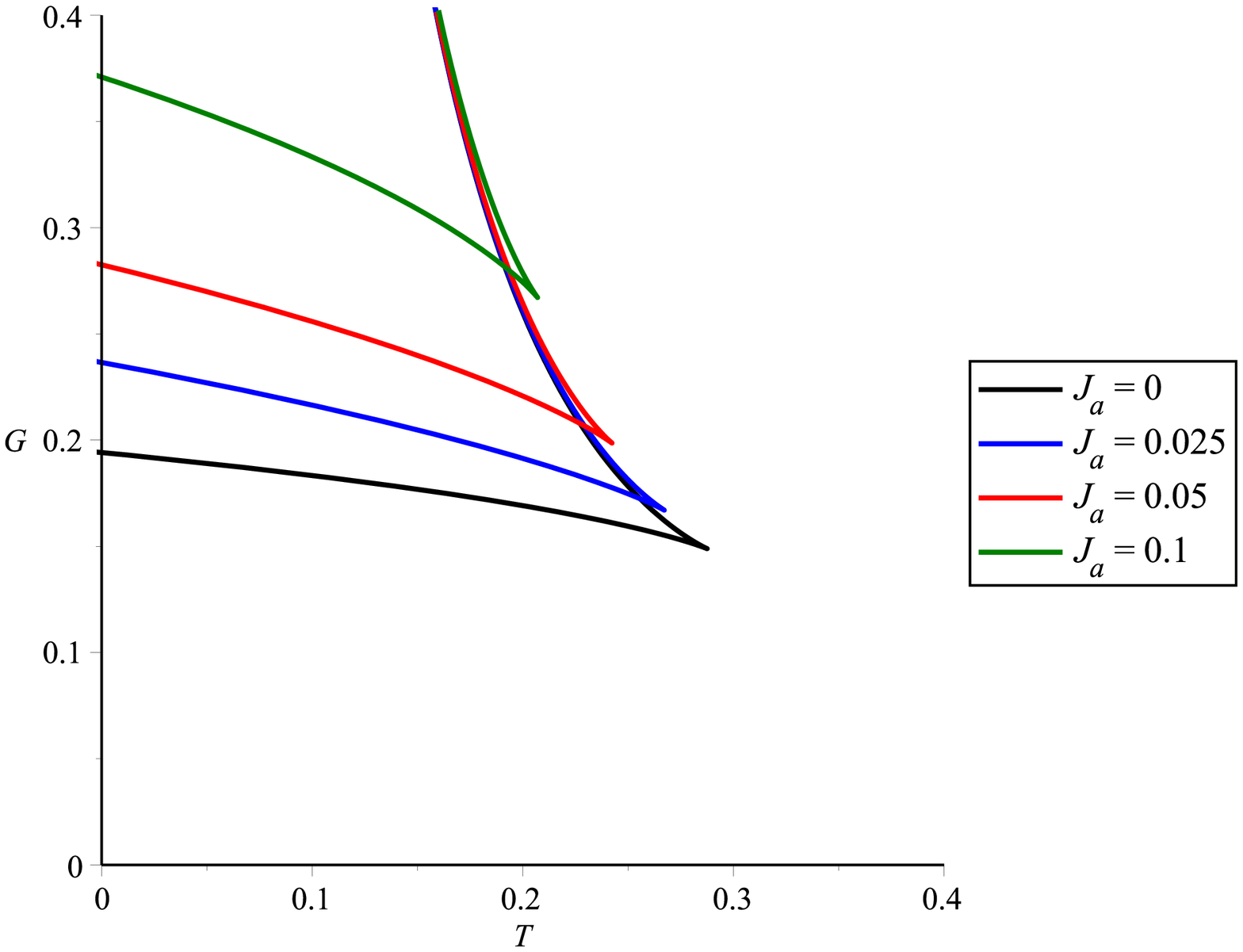}
	}
	\subfigure[$g=1$, $Q=0.04$ and $J_b=0.005$]{
		\includegraphics[width=0.46\textwidth]{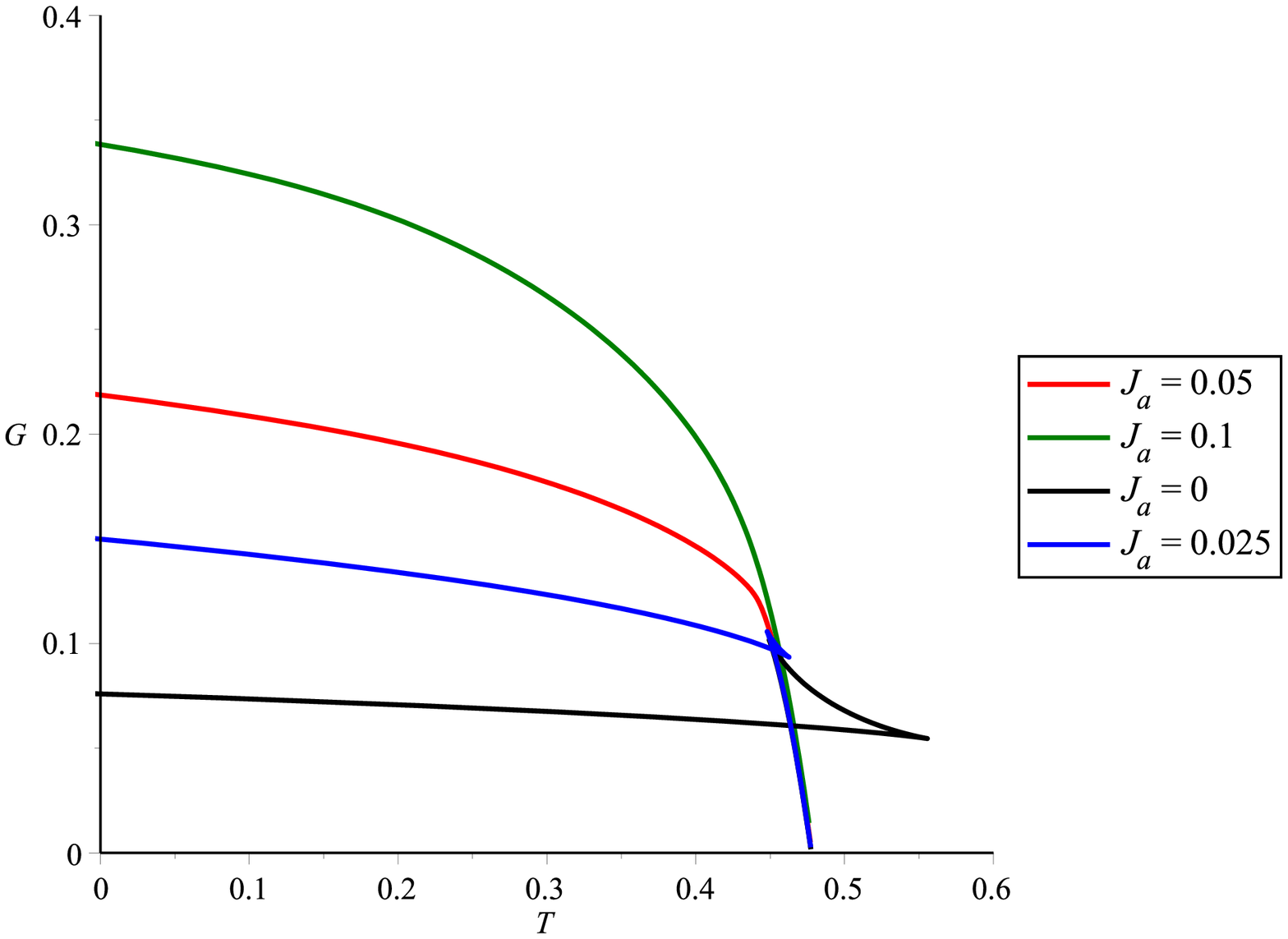}
	}
	
	\caption{Plots of the Gibbs free energy G(T) for fixed , $Q$, $J_a$, $J_b$ and $g$ (this implies $P={\rm const.}$)}
	\label{pic:PD_GT-plots}
\end{figure}

\begin{figure}[!h]
	\centering
	\includegraphics[width=0.5\textwidth]{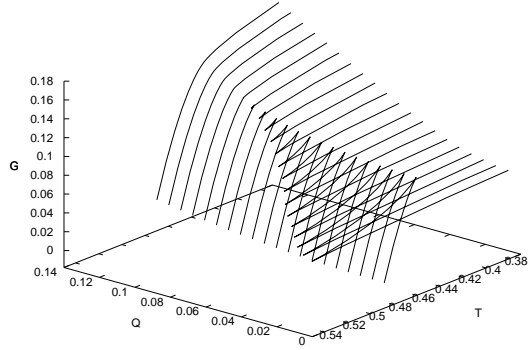}
	\caption{Three-dimensional plot of the Gibbs free energy $G(T,Q)$ for $g=1$, $J_a=0.005$ and $J_b=0.01$.}
	\label{pic:PD_GTQ-3D}
\end{figure}

In the canonical ensemble, meaning fixed charge and fixed angular momenta, one can compute the free energy. Since the mass $M$ can be interpreted as the enthalpy of the black hole, the Gibbs free energy $G(T, P, J_a, J_b,Q)$ is (see \cite{PD_Altamirano:2014tva})
\begin{equation}
	G=M-TS \, .
\end{equation}
For the rotating black hole in minimal five-dimensional gauged supergravity it is impossible to calculate $G$ analytically. Therefore we apply numerical methods as in the case of the state equation (section \ref{sec:PD_state-equation}). Plots $G(T)$ for various values of $g$, $Q$, $J_a$ and $J_b$ are depicted in figure \ref{pic:PD_GT-plots}. A three-dimensional plot of G versus T and Q (with fixed values of $g$, $J_a$ and $J_b$) is shown in figure \ref{pic:PD_GTQ-3D}. These plots confirm the results from section \ref{sec:PD_state-equation}. 

The graphs in figure \ref{pic:PD_GT-plots}(a), (c) and (e) have no cosmological constant, $g=0$. In the case $Q=J_a=J_b=0$ there is just one unstable black hole branch. If one or two angular momenta and/or the charge is present, a second branch of stable black holes emerges. 
\\

In the presence of a cosmological constant (figure \ref{pic:PD_GT-plots}(b), (d) and (f)), the so called swallowtails appear for certain parameters. The swallowtails illustrate the first order phase transition from small to large black holes. The branch coming from the left corresponds to the small black holes while the branch going down corresponds to the large black holes. In between the two discontinuity points, where the specific heat diverges, there is a branch of intermediate unstable black holes. If critical values of the parameters are reached (section \ref{sec:PD_state-equation}) the swallowtails disappear and a unique black hole phase remains.
In the three-dimensional plot of G versus T and Q (figure \ref{pic:PD_GTQ-3D}), it can be observed that the swallowtail vanishes as $Q$ reaches its critical value.
\\

\clearpage

\section{Thermodynamic stability}
\label{sec:PD_stability}
In this section we will analyse the thermodynamic stability of a rotating black hole in minimal five-dimensional gauged supergravity in the canonical ensemble. 
An important quantity is the specific heat at constant pressure $P$, charge $Q$ and angular momenta $J_a$, $J_b$
\begin{equation}
	C=C_{P,J_a,J_b,Q}= T \left. \frac{\partial S}{\partial T} \right| _{P,J_a,J_b,Q} \, .
\end{equation}
The ensure stability, the specific heat has to be positive. Additionally the isothermal moment of inertia 
\begin{equation}
	\epsilon = \epsilon _{T, P, Q} = \left. \frac{\partial J}{\partial \Omega} \right|_{T, P, Q}
\end{equation}
must be considered.

However, in this spacetime we have two angular momenta, so that the isothermal moment of inertia turns into a tensor
\begin{equation}
 \epsilon_{ij} = \left. \left( \frac{\partial J_i}{\partial \Omega_j} \right) \right|_{T, P, Q} \, ,
\end{equation}
with $i=a,b $ and $j=a,b$. For stability it is required that both of the eigenvalues of the moment of inertia tensor are positive \cite{PD_Monteiro:2009tc, PD_Dolan:2013yca, PD_Dolan:2014lea}.

A third condition for thermodynamic stability is the positivity of the adiabatic compressibility \cite{PD_Dolan:2014lea}
\begin{equation}
	k=k_{S,J_a,J_b,Q}= \left. -\frac{1}{V}\frac{\partial V}{\partial P} \right|_{S,J_a,J_b,Q} \, .
\end{equation}
The specific heat, the isothermal moment of inertia and the adiabatic compressibility can be found analytically in terms of the metric parameters $a$, $b$, $q$, $r_+$ and $g$, but the formulas tend to be extremely lengthy, so we will not display them here. 

To calculate the specific heat, the pressure P is held constant, which is equivalent to $g=\rm const.$. Equation \eqref{eqn:PD_charge} is solved to find an expression $q=q(Q, a, b, r_+, g)$, to be substituted into the equations \eqref{eqn:PD_entropy} and \eqref{eqn:PD_temperature} for entropy and temperature. Now the specific heat can be written as
\begin{equation}
 C=T \frac{ \frac{\partial S}{\partial a} + \frac{\partial S}{\partial b}\frac{\dd b}{\dd a} +  \frac{\partial S}{\partial r_+}\frac{\dd r_+}{\dd a} }{ \frac{\partial T}{\partial a} + \frac{\partial T}{\partial b}\frac{\dd b}{\dd a} +  \frac{\partial  T}{\partial r_+}\frac{\dd r_+}{\dd a}} \, .
\end{equation}
The derivatives $\frac{\dd b}{\dd a}$ and $\frac{\dd r_+}{\dd a}$ can be found from $J_a=\rm const.$ and $J_b=\rm const.$, since
\begin{align}
 \dd J_a &= \frac{\partial J_a}{\partial a}\dd a + \frac{\partial J_a}{\partial b}\dd b + \frac{\partial J_a}{\partial r_+}\dd r_+ = 0 \nonumber\\
 \dd J_b &= \frac{\partial J_b}{\partial a}\dd a + \frac{\partial J_b}{\partial b}\dd b + \frac{\partial J_b}{\partial r_+}\dd r_+ = 0 \, .
\end{align}

To find the isothermal moment of inertia tensor (see also \cite{PD_Dolan:2014lea}), we  fix $g$ and substitute $q=q(Q, a, b, r_+, g)$ into the angular momenta (equation \eqref{eqn:PD_angularmomenta}) the angular velocities (equation \eqref{eqn:PD_omega}) and the temperature (equation \eqref{eqn:PD_temperature}). From the condition $T=\rm const.$ and therefore
\begin{equation}
 \dd T = \frac{\partial T}{\partial a}\dd a + \frac{\partial T}{\partial b}\dd b + \frac{\partial T}{\partial r_+}\dd r_+ = 0
\end{equation}
we get the partial derivatives $\frac{\partial r_+}{\partial a}$ and $\frac{\partial r_+}{\partial b}$. These can be used to calculate
\begin{equation}
  \left. \frac{\partial J_i}{\partial j} \right|_{T, P, Q} \quad {\rm and} \quad \left. \frac{\partial \Omega_i}{\partial j} \right|_{T, P, Q}
\end{equation}
with $i=a,b $ and $j=a,b$. $\frac{\partial \Omega_i}{\partial j}$ can be inverted as a matrix and finally the moment of inertia tensor is
\begin{equation}
	 \epsilon_{ij} = \left. \frac{\partial J_i}{\partial k} \frac{\partial k}{\partial \Omega_j}  \right|_{T, P, Q} \, .
\end{equation}
$\epsilon_{ij}$ has two eigenvalues $\lambda_1$ and $\lambda_2$, which have to be positive for thermodynamic stability.

The adiabatic compressibility can be calculated analogously to the specific heat. Although we could not find a rigorous proof, the adiabatic compressibility was positive in all parameter ranges we tested (as long as $a^2<1/g^2$ and $b^2<1/g^2$), so we will not consider it in the following analysis. For the asymptotically AdS Kerr spacetime with two independent rotation parameters it is already known that $k$ is positive \cite{PD_Dolan:2013dga}.
\\

\begin{figure}
	\centering
	\subfigure[Sign of $\lambda_1$ for $Q=0.25$]{
		\includegraphics[width=0.31\textwidth]{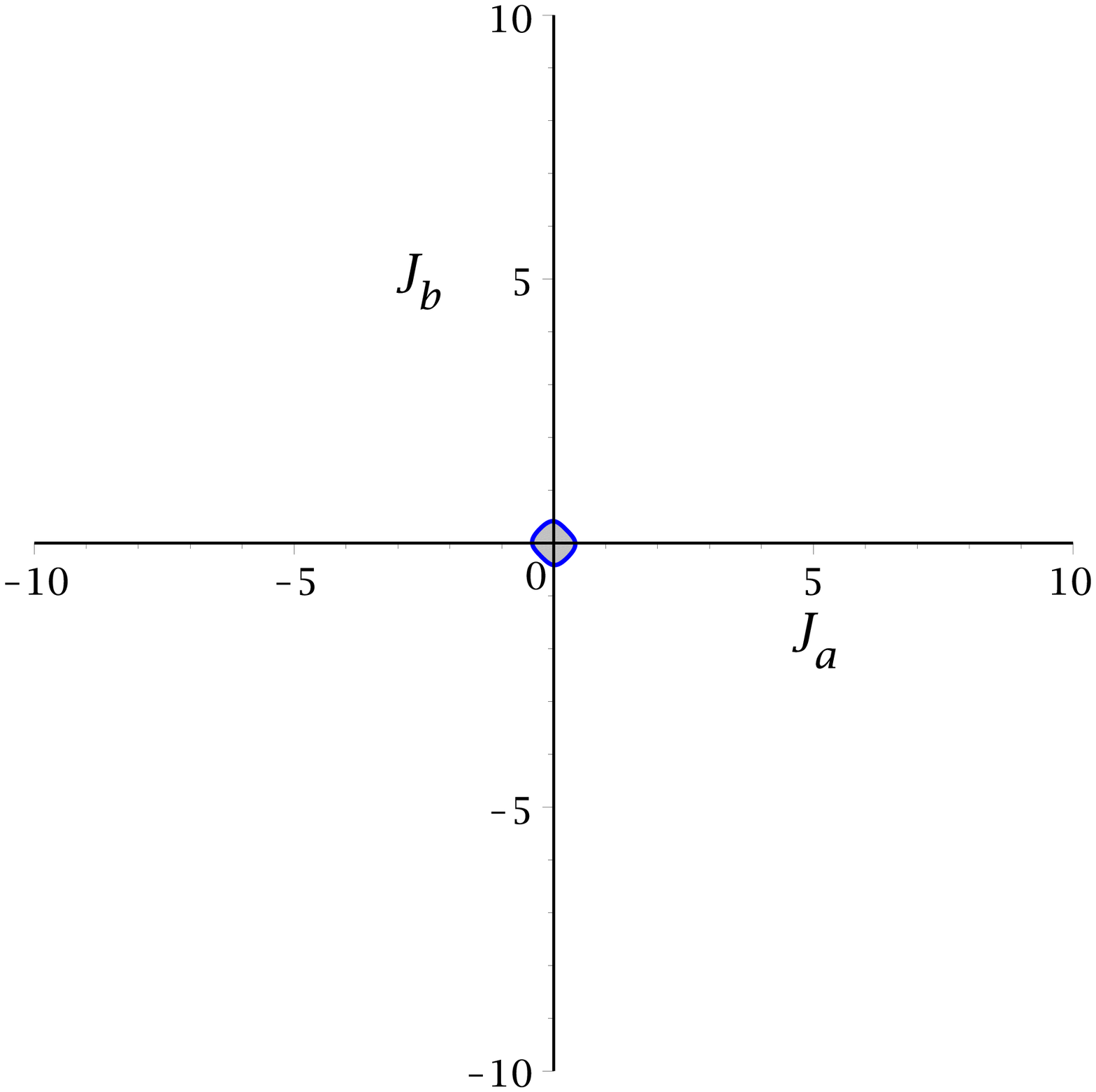}
	}
	\subfigure[Sign of $\lambda_2$ for $Q=0.25$]{
		\includegraphics[width=0.31\textwidth]{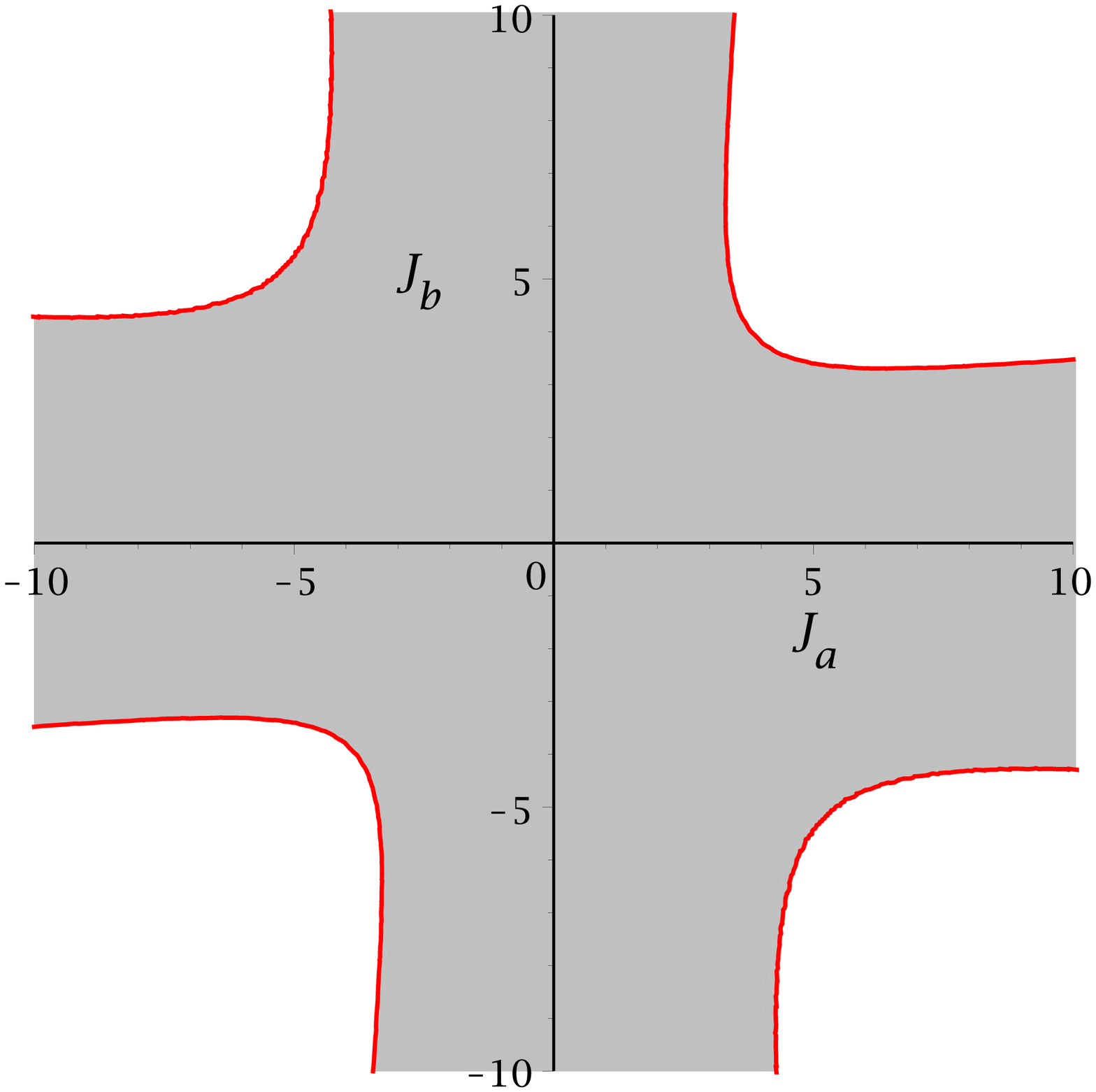}
	}
	\subfigure[Sign of $C_J$ for $Q=0.25$]{
		\includegraphics[width=0.31\textwidth]{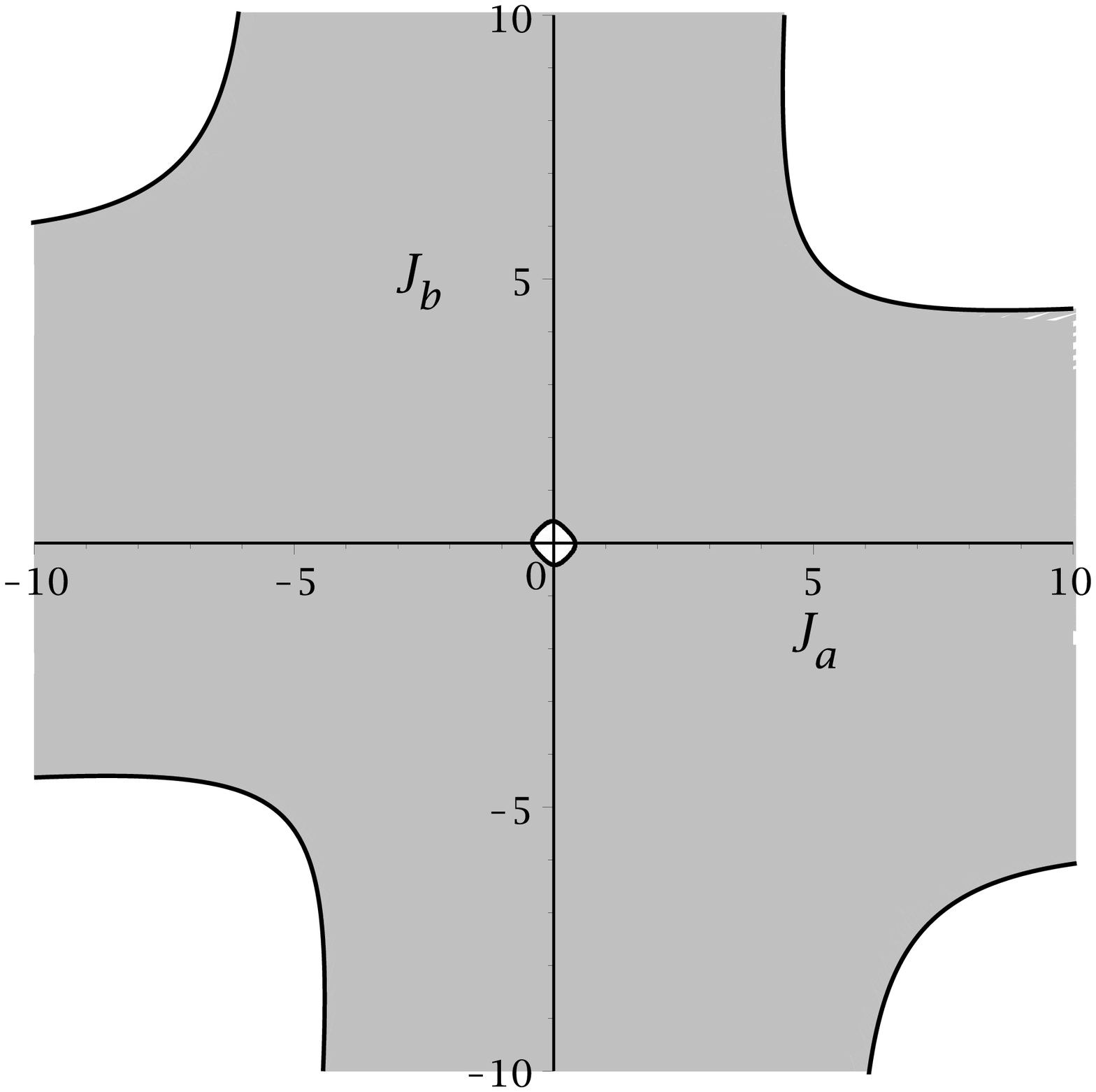}
	}
	
	\subfigure[Sign of $\lambda_1$ for $Q=0.75$]{
		\includegraphics[width=0.31\textwidth]{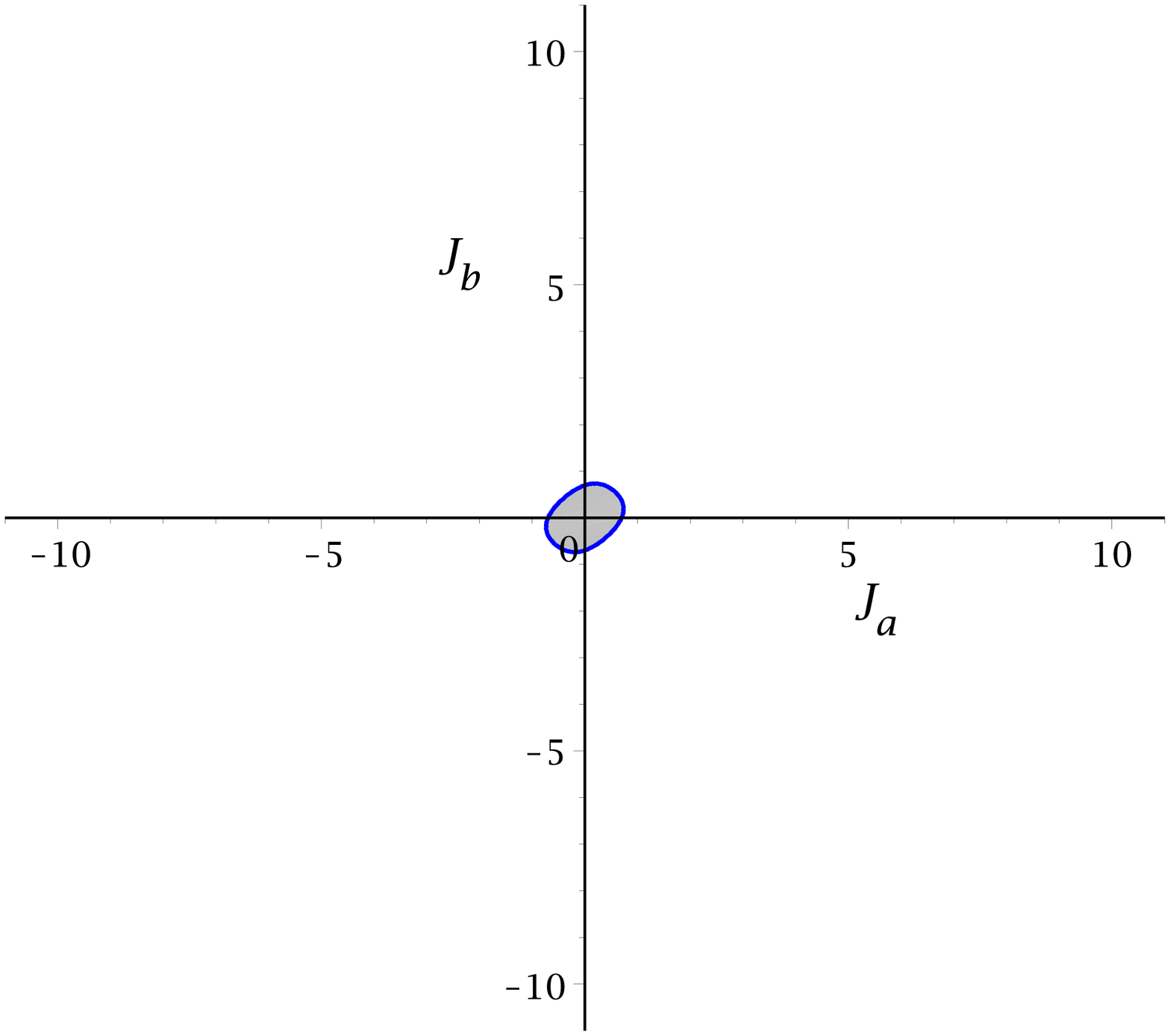}
	}
	\subfigure[Sign of $\lambda_2$ for $Q=0.75$]{
		\includegraphics[width=0.31\textwidth]{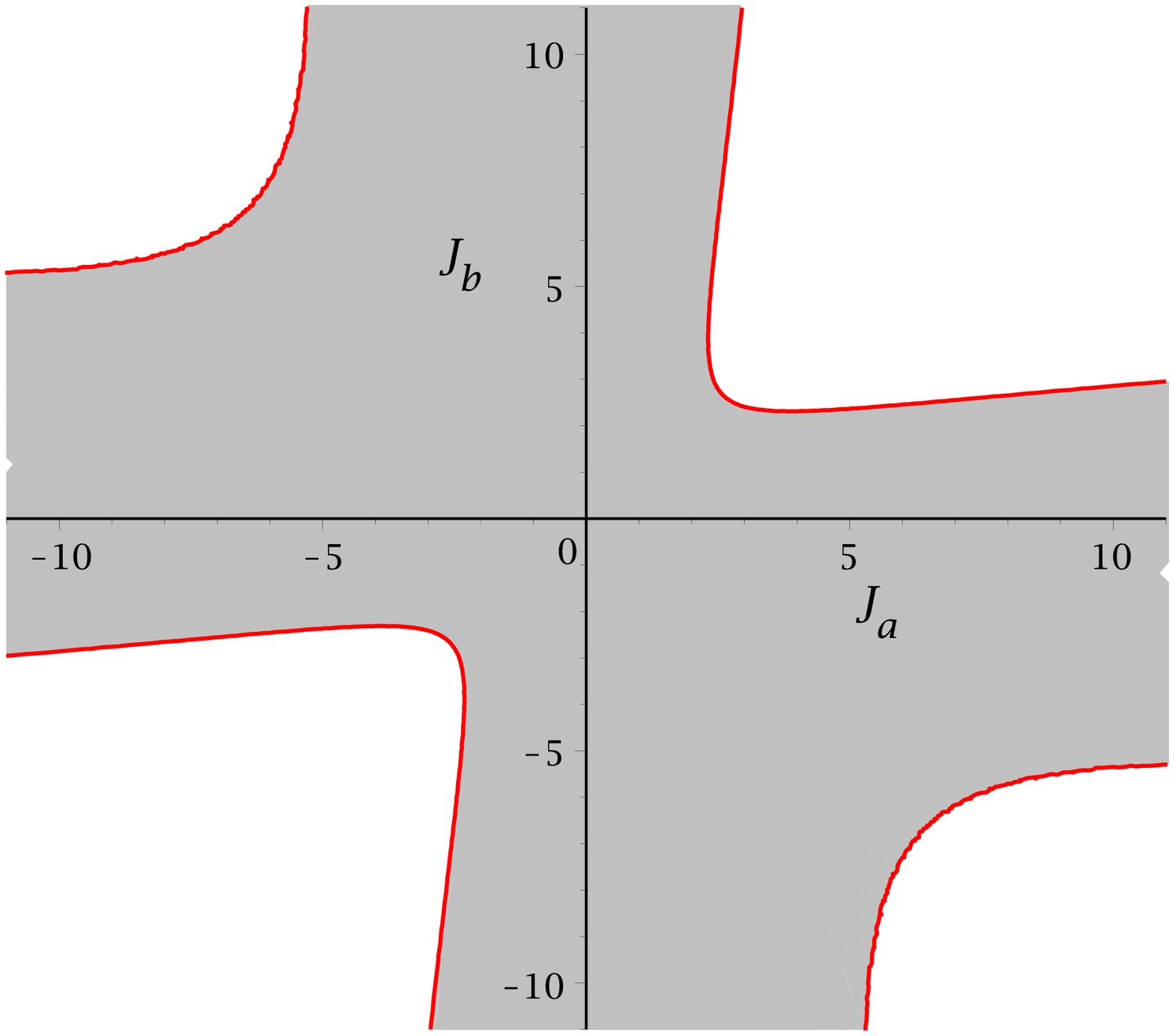}
	}
	\subfigure[Sign of $C_J$ for $Q=0.75$]{
		\includegraphics[width=0.31\textwidth]{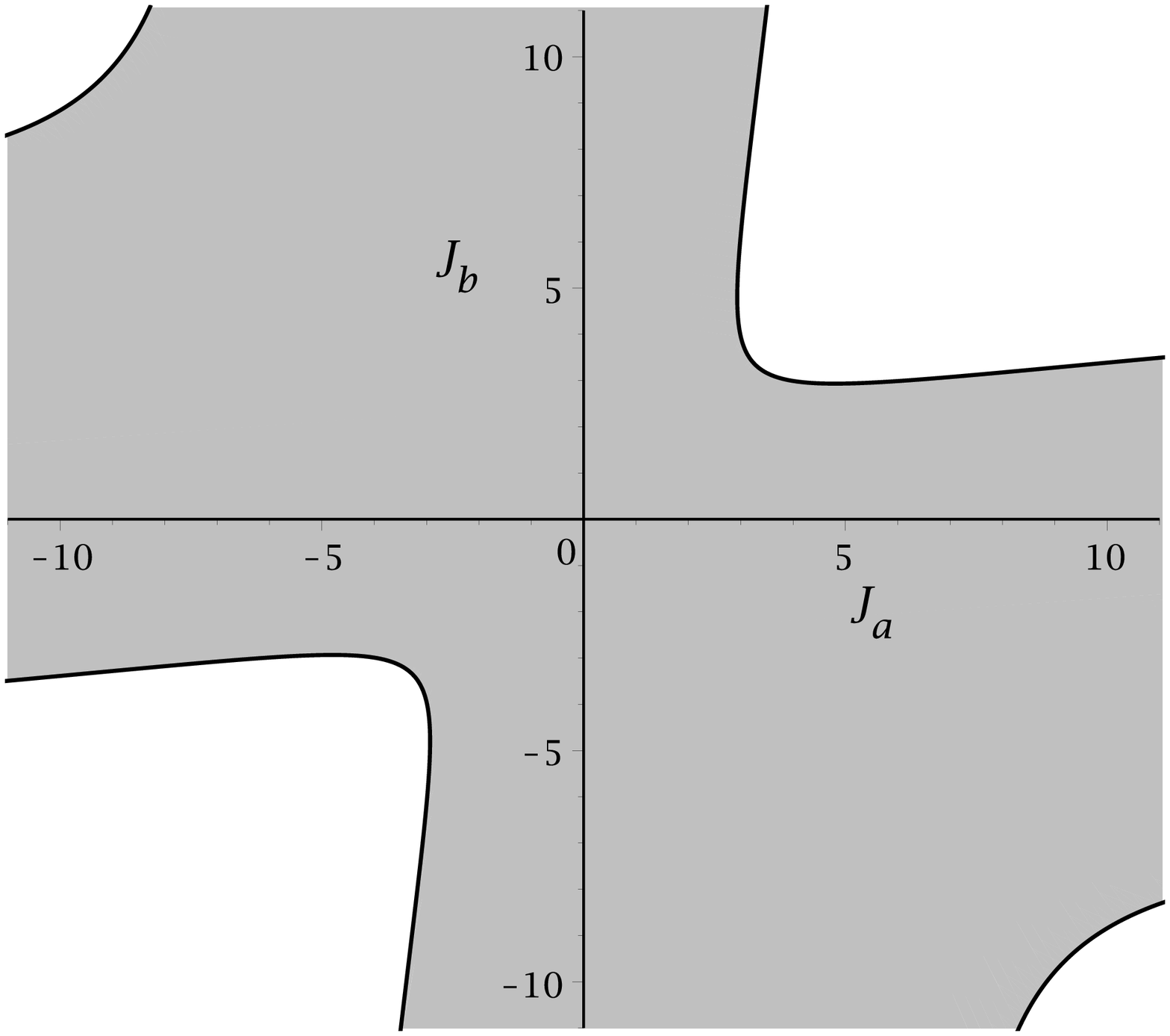}
	}
	
	\subfigure[Sign of $\lambda_1$ for $Q=1.25$]{
		\includegraphics[width=0.31\textwidth]{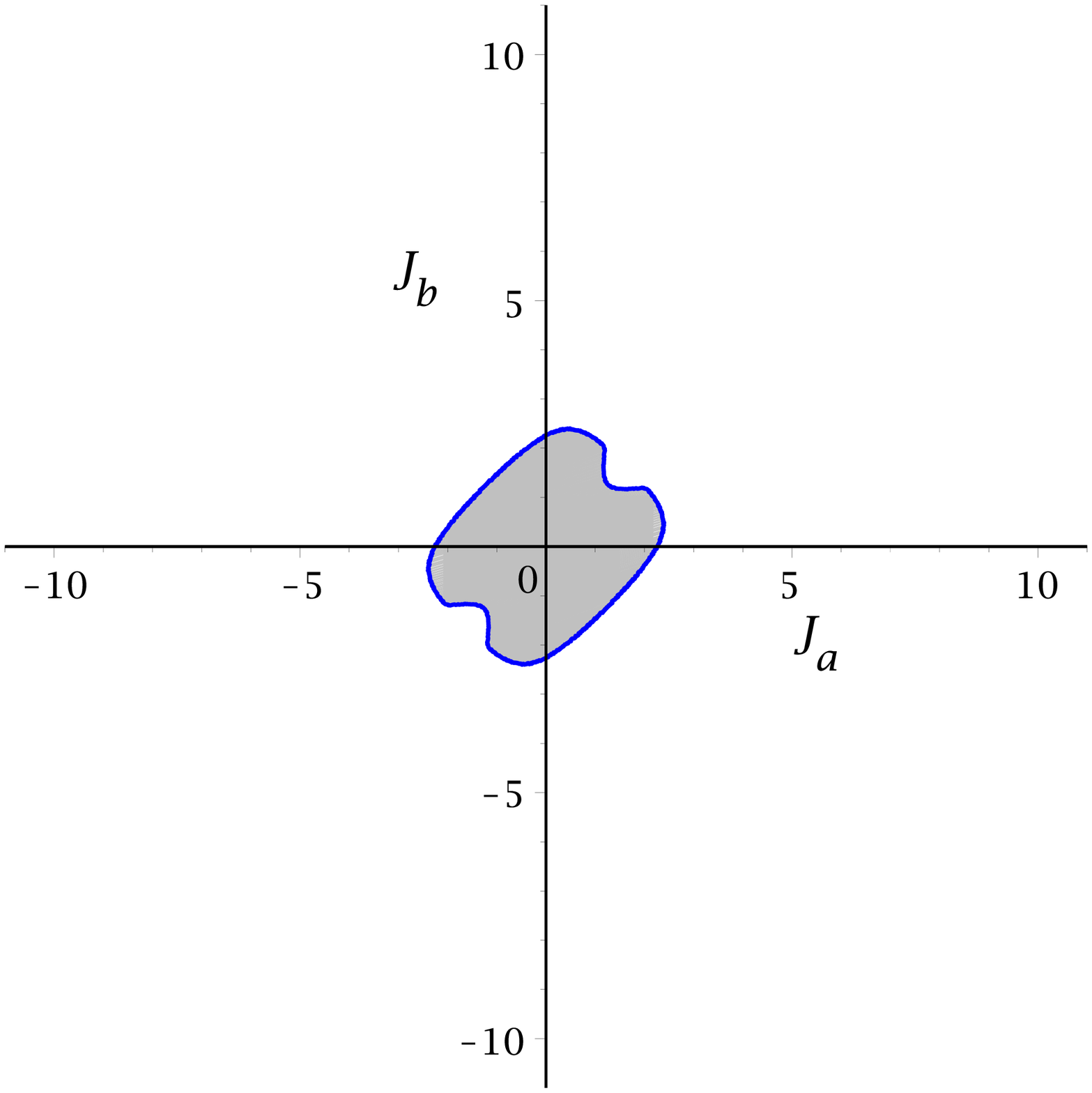}
	}
	\subfigure[Sign of $\lambda_2$ for $Q=1.25$]{
		\includegraphics[width=0.31\textwidth]{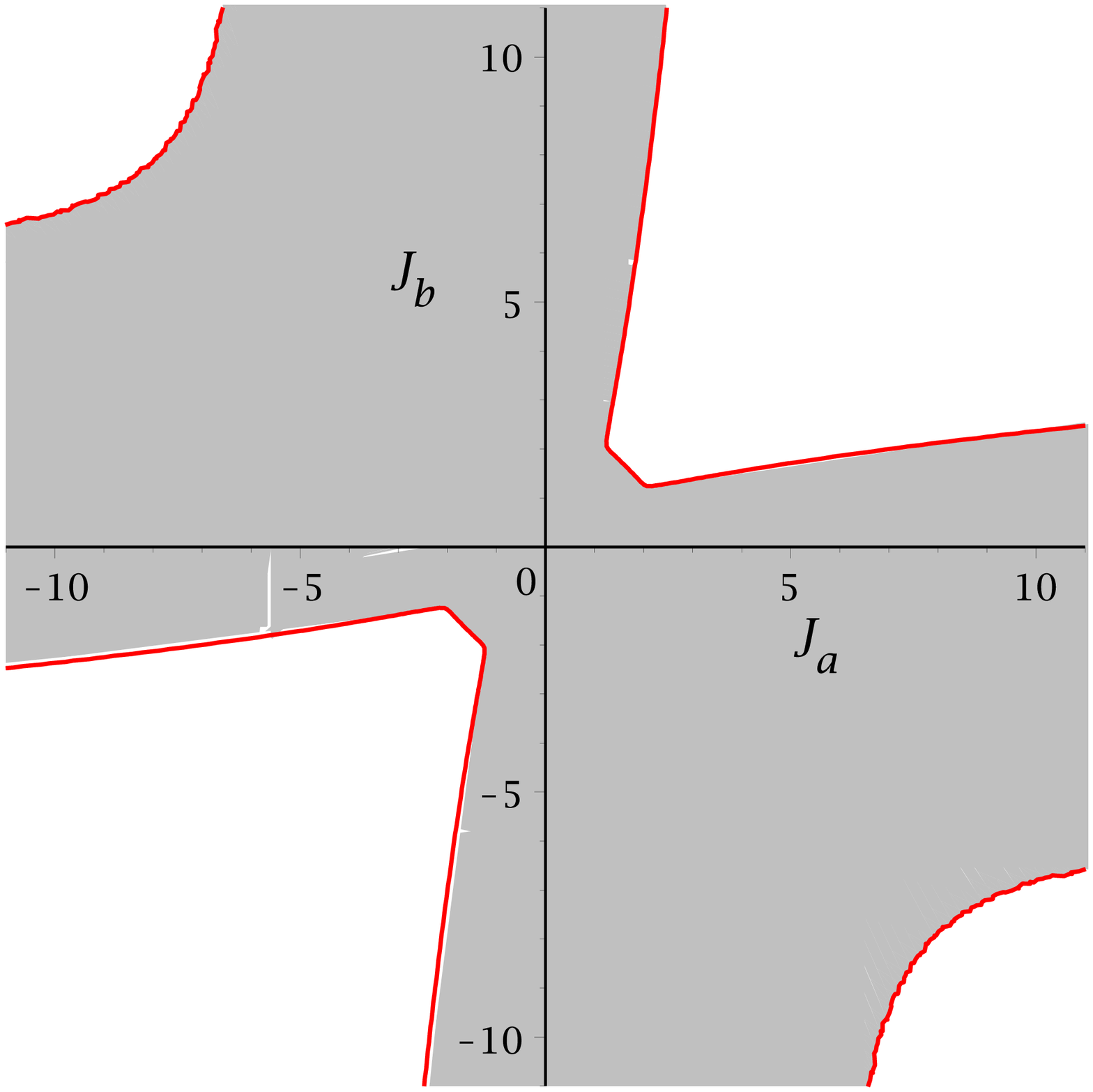}
	}
	\subfigure[Sign of $C_J$ for $Q=1.25$]{
		\includegraphics[width=0.31\textwidth]{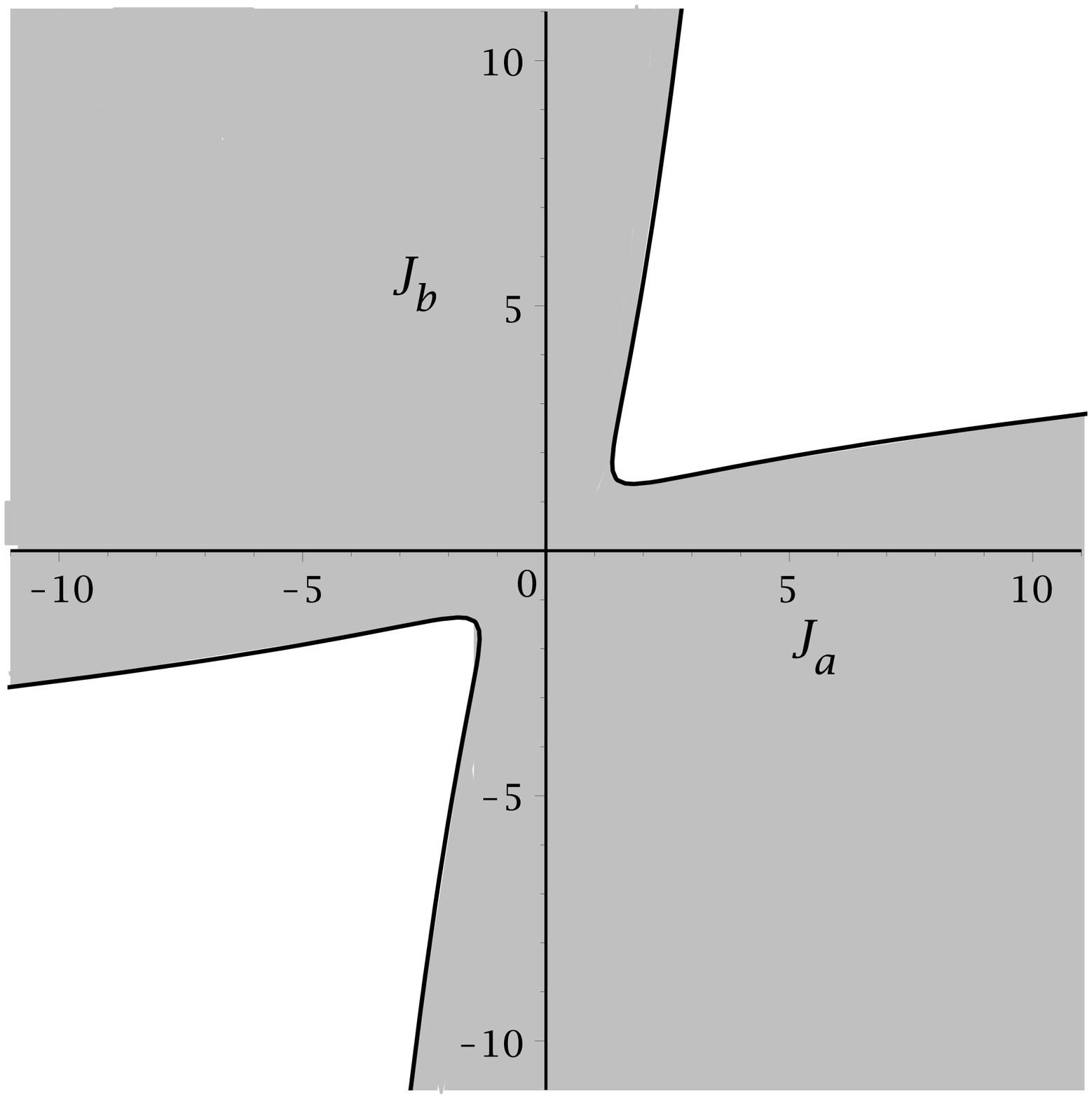}
	}
	
	\subfigure[Sign of $\lambda_1$ for $Q=1.5$]{
		\includegraphics[width=0.31\textwidth]{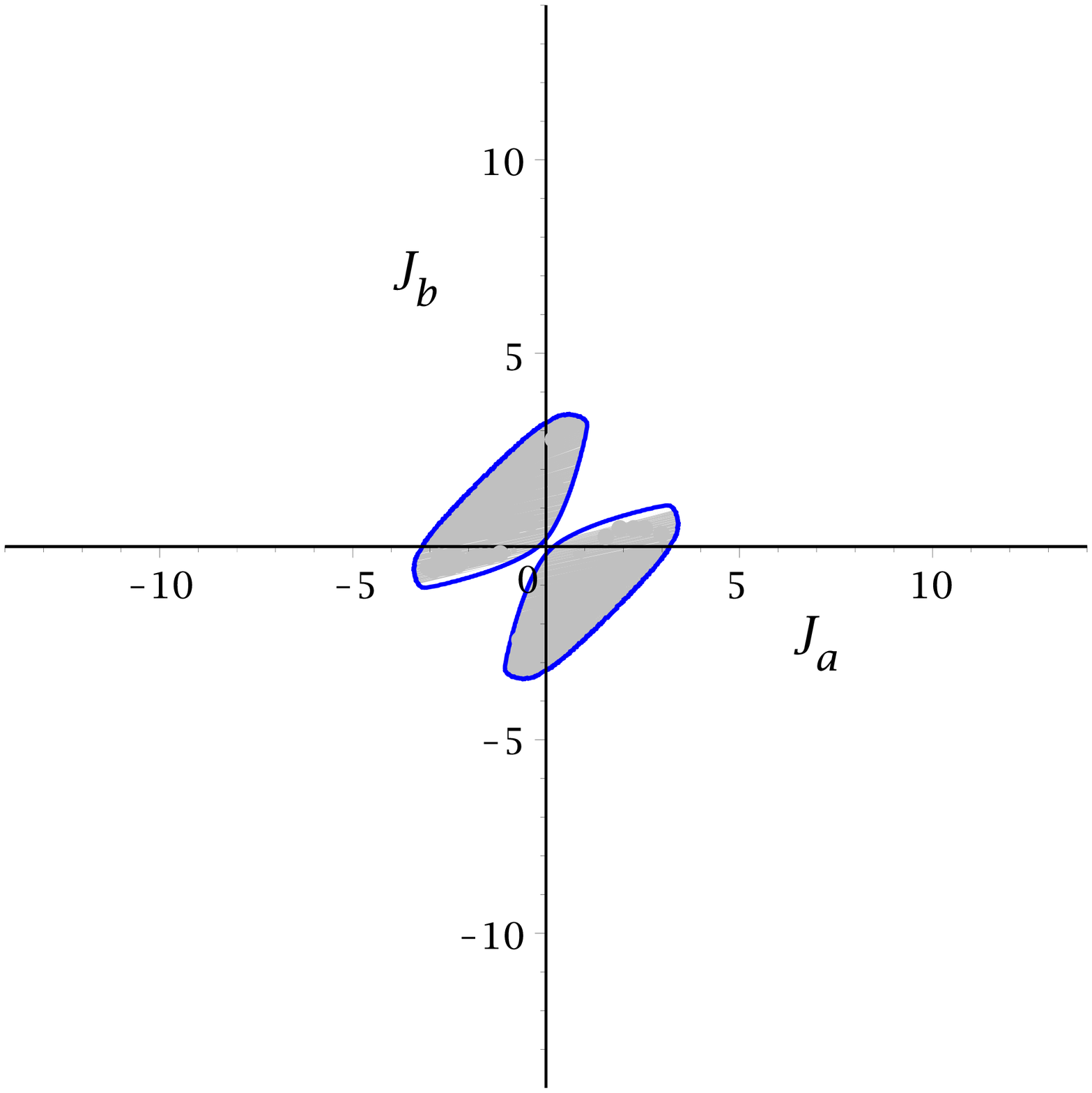}
	}
	\subfigure[Sign of $\lambda_2$ for $Q=1.5$]{
		\includegraphics[width=0.31\textwidth]{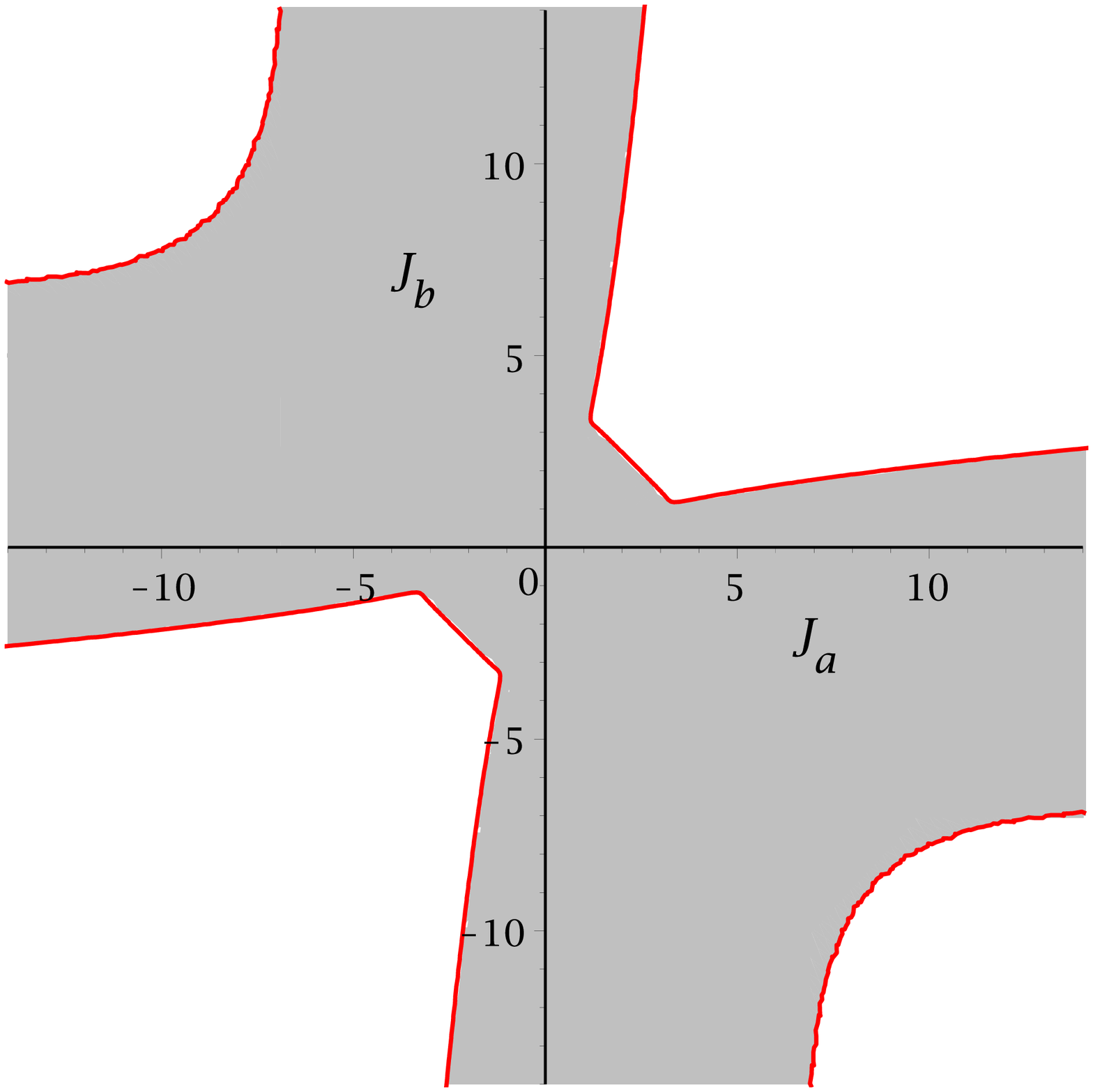}
	}
	\subfigure[Sign of $C_J$ for $Q=1.5$]{
		\includegraphics[width=0.31\textwidth]{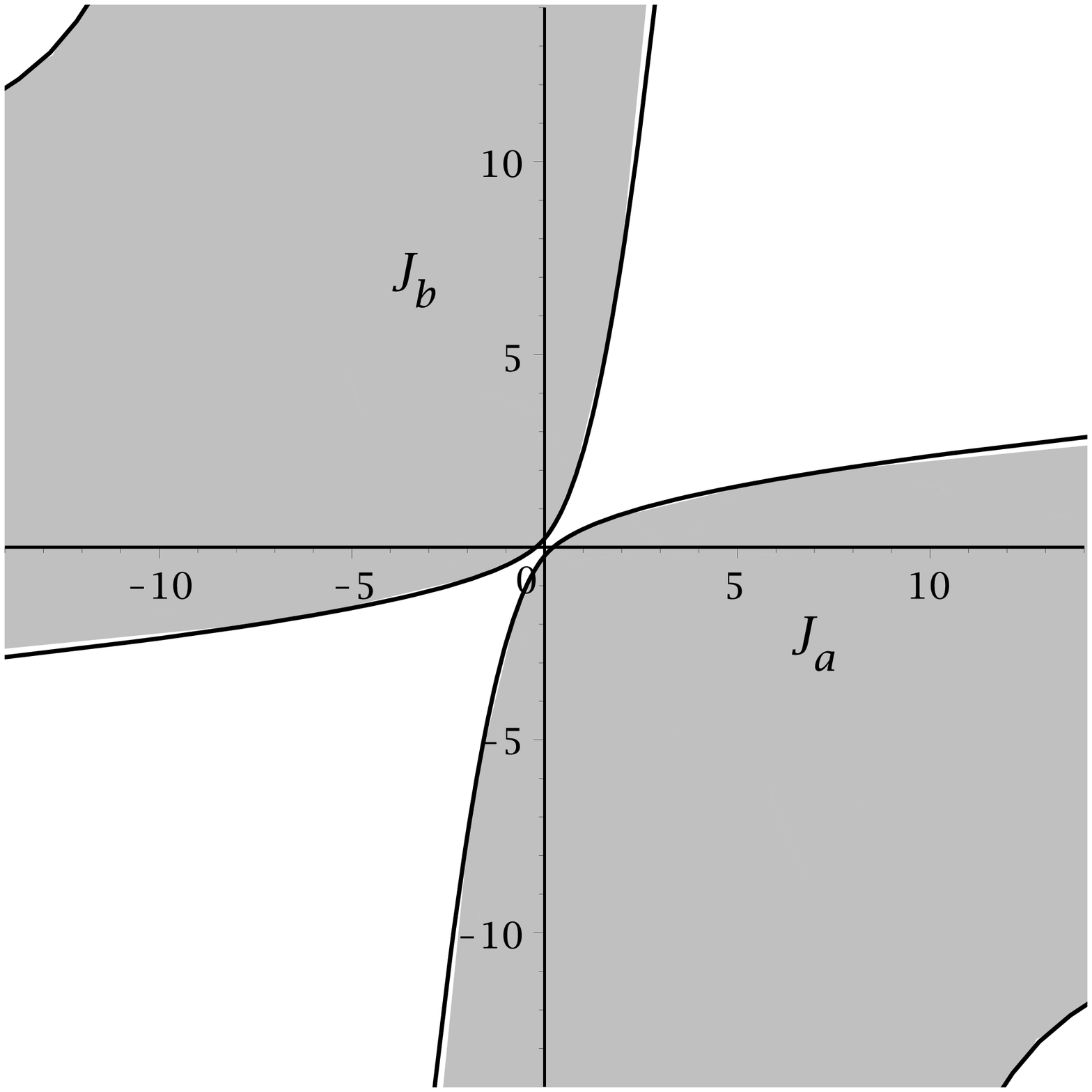}
	}
	\caption{Sign of the specific heat $C$ at constant angular momenta  and the two eigenvalues $\lambda_1$, $\lambda_2$ of the isothermal moment of inertia tensor at $r_+=1$ and $g=0.3$. The black hole is thermodynamically stable if $\lambda_1$, $\lambda_2$ and $C$ are all positive at the same time.}
	\label{pic:PD_stability}
\end{figure}

Figure \ref{pic:PD_stability} shows the sign of the specific heat and the two eigenvalues of the isothermal moment of inertia tensor for different values of the charge $Q$. The parameters for the cosmological constant and the event horizon are fixed to $g=0.3$ and $r_+=1$. This leaves two free parameters $a$, $b$, but since we want to display the sign of $C$, $\lambda_1$ and $\lambda_2$ in dependence of the angular momenta, we perform a ``coordinate change'' in the plots from $a$, $b$ to $J_a$, $J_b$ using internal routines of the computer algebra system Maple and equation \eqref{eqn:PD_angularmomenta}. The grey regions correspond to positive $C$, $\lambda_1$ and $\lambda_2$ while the white regions correspond to negative values and therefore instability.

The system is thermodynamically stable if $C$, $\lambda_1$ and $\lambda_2$ are positive at the same time. This is possible for many cases, but not for all. If $g=0$ and $Q=0$, then the black hole is unstable.

For $|g|>0$ and/or $|Q|>0$, a region of stability appears, whose shape is mostly determined by the eigenvalue $\lambda_1$. So an uncharged black hole with a cosmological constant is stable for certain $J_a$ and $J_b$, but also a charged asymptotically flat black hole is stable. The stability region grows with increasing $Q$ until it separates into two regions, leaving an instability region around $J_a=J_b=0$ in between.

If the black hole is uncharged, $Q=0$, the plots of the sign of $C$, $\lambda_1$ and $\lambda_2$ are symmetric to the $J_a$-axis and to the $J_b$-axis. In the case of a charged black hole, the plots are symmetric with respect to the diagonals.

\section{Conclusion}

In this paper we analysed the thermodynamics of a rotating black hole in minimal five-dimensional gauged supergravity. First a brief review of the metric and its thermodynamic quantities was given. Due to the presence of the cosmological constant there is an extra term $V\dd P$ in the first law of thermodynamics, where $P$ is a pressure and $V$ is ab effective volume inside the event horizon. The mass $M$ corresponds to the total gravitational enthalpy of the system.
\\

Then the state equation $T(S, J_a, J_b, Q)$ was analysed for fixed charge $Q$, fixed angular momenta $J_a$, $J_b$ and fixed pressure $P$. For $g=0$, the typical Schwarzschild, Reissner-Nordstr\"om and Kerr behaviour can be observed. For $g\neq 0$, depending on the parameters, a first order phase transition from small black holes to large black holes occurs, resembling the Van der Waals liquid/gas phase transition. After critical values of $g$, $J_a$, $J_b$ and $Q$ are reached, only one black hole phase is possible. This indicates a second order phase transition. The critical values form a circle in the $J_a$-$J_b$-plane, inside the circle two phases are possible and outside the circle just one phase exists. The circle shrinks with increasing $g$ and $Q$, so that for large $g$ or $Q$ there is only a unique phase of black holes.

The study of the free energy $G=M-TS$ (which is the Gibbs free energy since $M$ is the enthalpy), confirms these results. In the case of two black hole phases, the famous swallowtails are found in the $G$-$T$-plots.
\\

In the last section the thermodynamic stability was considered. To guarantee stability three conditions must be fulfilled \cite{PD_Dolan:2014lea}: 
\begin{itemize}
	\item $\displaystyle	C= T \left. \frac{\partial S}{\partial T} \right| _{P,J_a,J_b,Q} >0 $ 
	\item Both eigenvalues of $\displaystyle \epsilon_{ij} = \left. \left( \frac{\partial J_i}{\partial \Omega_j} \right) \right|_{T, P, Q}$ (with $i=a,b$ and $j=a,b$) have to be positive.
	\item $\displaystyle k= \left. -\frac{1}{V}\frac{\partial V}{\partial P} \right|_{S,J_a,J_b,Q} > 0$
\end{itemize}
Using plots of the sign of $C$ and the eigenvalues of $\epsilon_{ij}$, it was shown, that the rotating black hole in minimal five-dimensional gauged supergravity is thermodynamically stable in certain parameter ranges.
\\

In five dimensions other than spherical event horizon topologies are possible. It would be interesting to compare the results of this paper to non-spherical black holes in AdS, like black rings or black saturns. Thin black rings and also thin black saturns have already been constructed \cite{PD_Caldarelli:2008pz} using the blackfold approach \cite{PD_Emparan:2009at}. The thermodynamics of the singly spinning thin black ring were studied in \cite{PD_Altamirano:2014tva}. The specific heat of the thin AdS black ring is always negative, making the black ring thermodynamically unstable. However, this may be different in the parameter range where the blackfold approach is no longer applicable. Although arguments have been presented that supersymmetric AdS black rings do not exist (at least not without a conical singularity) \cite{PD_Kunduri:2006uh, PD_Kunduri:2007qy}, there might be a non-supersymmetric AdS black ring solution waiting to be found  \cite{PD_Caldarelli:2008pz}. There is still no analytical solution, but very recently a numerical solution of an AdS black ring was presented in \cite{PD_Figueras:2014dta}.

\section{Acknowledgements}
We would like to thank Jutta Kunz and Niels Obers for fruitful discussions. Furthermore S.G. is grateful to the Niels Bohr Institute in Copenhagen, where part of this was performed.
S.G. gratefully acknowledges support by the DFG within the Research Training Group 1620 ``Models of Gravity''.

\clearpage

\bibliographystyle{unsrt}

\end{document}